\documentclass[fleqn,usenatbib]{mnras}

\usepackage{newtxtext,newtxmath}

\usepackage[T1]{fontenc}
\usepackage{ae,aecompl}

\usepackage{graphicx}	
\usepackage{amsmath}	
\usepackage{amssymb}	
\usepackage{textcomp}
\usepackage{booktabs}
\usepackage{pdflscape}
\usepackage{xcolor}
\usepackage{booktabs}
\usepackage{longtable}

\title[Dust Obscured Star Formation in CLJ1449]{Revealing Dust Obscured Star Formation in CLJ1449+0856, a Cluster at z=2 }

\author[Connor M. A. Smith et al.]{
C. M. A. Smith,$^{1}$\thanks{E-mail: Connor.Smith@astro.cf.ac.uk}
W. K. Gear,$^{1}$
M. W. L. Smith,$^{1}$
A. Papageorgiou,$^{1}$ 
\newauthor \, S. A. Eales$^{1}$
\\
$^{1}$School of Physics and Astronomy, Cardiff University, Queens Building, The Parade, Cardiff, CF24 3AA, UK\\
}

\date{Accepted XXX. Received YYY; in original form ZZZ}

\pubyear{2019}

\begin{document}

\label{firstpage}
\pagerange{\pageref{firstpage}--\pageref{lastpage}}
\maketitle

\defcitealias{2018Coogan}{C18}
\defcitealias{2018Strazzullo}{S18}

\begin{abstract}
We present SCUBA-2 450\,$\mu$m and 850\,$\mu$m data of the mature redshift 2 cluster CLJ1449. We combine this with archival \textit{Herschel} data to explore the star forming properties of CLJ1449. Using high resolution ALMA and JVLA data we identify potentially confused galaxies, and use the Bayesian inference tool XID+ to estimate fluxes for them. Using archival optical and near infrared data with the energy-balance code \textsc{CIGALE} we calculate star formation rates, and stellar masses for all our cluster members, and find the star formation rate varies between 20-1600\,M$_{\odot}$yr$^{-1}$ over the entire 3\,Mpc radial range. The central 0.5\,Mpc region itself has a total star formation rate of 800\,$\pm$\,200\,M$_{\odot}$yr$^{-1}$, which corresponds to a star formation rate density of (1.2\,$\pm$\,0.3)\,$\times$10$^{4}$\,M$_{\odot}$yr$^{-1}$Mpc$^{-3}$, which is approximately five orders of magnitude greater than expected field values. When comparing this cluster to those at lower redshifts we find that there is an increase in star formation rate per unit volume towards the centre of the cluster. This indicates that there is indeed a reversal in the star formation/density relation in CLJ1449. Based on the radial star-formation rate density profile, we see evidence for an elevation in the star formation rate density, even out to radii of 3\,Mpc. At these radii the elevation could be an order of magnitude greater than field values, but the exact number cannot be determined due to ambiguity in the redshift associations. If this is the case it would imply that this cluster is still accreting material which is possibly interacting and undergoing vigorous star-formation.

\end{abstract}




\section{Introduction}

Ever since the appearance of the `Lilly-Madau' plots of the 1990's (\citealt{1996Lilly}, \citealt{1996Madau}), which showed the change of Star Formation Rate (SFR) density against time (or redshift), the evolution of the SFR in the universe has been one of the main focuses in observational cosmology. Large samples of `sub-millimetre galaxies' (SMGs) produced by ground-based telescopes such as the James Clerk Maxwell telescope (JCMT), and most recently the \textit{Herschel} satellite (\citealt{2010Pilbratt}) have played a central, often leading role in these studies, especially when tracing the obscured star formation. As well as the universal evolution of the SFR density however, its variation with environment is also a crucial element of the process of understanding the evolution of galaxies in the Universe.

It is well known that the environment a galaxy resides in plays a major role in how it evolves. Tracers of galaxy evolution, such as morphology, colour and star formation, have all shown that as galaxy density increases, 'red and dead'  elliptical galaxies become dominant (e.g., \citealt{1980Dressler}, \citealt{2005Smith}, \citealt{2006Baldry}, \citealt{2006Haines}, \citealt{2002Lewis}, \citealt{2003Gomez}). This is in stark contrast to what is found in low density  environments, where blue, star forming spiral galaxies dominate.

At some point in cosmic time, the ellipticals that now dominate clusters in the local universe must have been actively forming stars. Even though in the local universe, the bulk of star formation is contributed by field galaxies, as we go to higher redshifts clusters become far more important, and must contribute significantly at the peak of star formation, approximately at redshift 2 (e.g. \citealt{2014Madau}). Indeed it has been shown (e.g. \citealt{2011Magnelli}, \citealt{2013Haines}) that both luminous infrared galaxies (LIRGs) and ultra luminous infrared galaxies (ULIRGs), highly star forming galaxies are virtually absent in cluster environments up until $z\sim 0.5$. Recent theoretical work has also predicted that in the early universe, it was in the over-dense proto-cluster regions where the majority of star formation took place (\citealt{2013Chiang,2017Chiang}). 

The evolution of star formation has been well detailed in clusters up to a redshift $z\sim1$, with the mass normalised SFR (SFR/Cluster Mass) being proportional to $(1+z)^{n}$, with $n$ being somewhere between 2-7 (e.g. \citealt{2004Cowie}, \citealt{2004Kodama}, \citealt{2006Geach}, \citealt{2008Saintonge}). However beyond z$\sim$1.5 the number of studied systems decreases rapidly, with only a handful being studied in detail. Even though  numbers are  low it does seem that these high redshift systems still follow this power law (e.g. \citealt{2014Santos,2015Santos}, \citealt{2014Smail},  \citealt{2015Ma}, \citealt{2015Webb}, \citealt{2016Casey}).

At increasing redshifts the obscuring effects of dust become more prominent and hence surveys using longer wavelengths such as infrared (IR) and sub-millimeter (sub-mm) have to be performed (\citealt{2005Lefloch}). Observations with both the \textit{Infrared space observatory} (ISO) and \textit{Spitzer} have shown that most of the SF can only be traced with IR wavelengths (e.g., \citealt{2002Duc}, \citealt{2005Metcalfe}, \citealt{2006Geach}, \citealt{2008Marcillac}), with \cite{2005Lefloch} showing that LIRGs contribute 70\% of the energy density at redshift one.

The first evidence for a reversal of the SFR-density relation was reported by \cite{2007Elbaz} who showed that for field galaxies between 0.8$<z<$1.2, the star formation rate increased with galaxy density, which is in stark contrast to what is seen at low redshift. The first work on clusters at high redshift was by \cite{2010Hayashi} and \cite{2010Tran}, who looked at $z\sim$1.4 and $z\sim$1.6 respectively. Again they found evidence for an increase in the star formation with increasing galaxy density. On the other hand \cite{2013Santos} found no evidence for a reversal of the SFR-density relation in a cluster at $z=1.4$.

Many of these studies used both near and mid IR (NIR, MIR) instruments such as \textit{Spitzer}, but at increasing redshift these wavelengths becomes a far less reliable tracer for star formation (e.g., \citealt{2007Calzetti}, \citealt{2009Hainline}). Instead far-infrared (FIR) and sub-mm tracers have to be employed. Facilities such as \textit{Herschel} and the  SCUBA-2 camera (\citealt{SCUBA2}), on the JCMT have allowed us to further understand the luminous properties of distant clusters. The ability to retain a high sensitivity over wide areas have enabled the study of entire clusters to investigate the SF-density relation. Such FIR/Submm studies of high redshift clusters have all shown significant populations of LIRGs and ULIRGs (\citealt{2014Santos,2015Santos}, \citealt{2014Smail}, \citealt{2015Ma}). This all leads to strong evidence that at these redshifts environmental effects have not yet taken hold and galaxies in clusters can still readily form stars. The statistics are still small, but with the advent of new techniques and facilities we are discovering more high redshift mature clusters (i.e. fully formed, virialized and emitting X-rays) and proto-clusters (i.e. clusters that are still in the process of forming), allowing for more robust determination of whether there is a reversal at high redshift (e.g. \citealt{2014Andereon}, \citealt{2014Clements}, 
\citealt{2015Bleem}, \citealt{2016Alexander}, \citealt{2016Franck}, \citealt{2017Cai}, \citealt{2017Daddi}, \citealt{2017Mantz}, \citealt{2017Umehata}, \citealt{2018Casasola}, \citealt{2018Lewis}, \citealt{2018Oteo}, \citealt{2018Zeballos})

CL J1449+0856 (hereafter CLJ1449, RA= 222.3083 Dec=8.9392) is one of the highest redshift, fully virialized, mature X-ray emitting clusters known. The cluster was first identified as an over-density of red galaxies during \textit{Spitzer} observations of the so called `Daddi fields' (\citealt{2000Daddi}). Optical, NIR and X-ray follow ups confirmed this was indeed fully virialised (\citealt{2006Kong}, \citealt{2011Gobat}). Follow up spectroscopy confirmed the redshift of the cluster to be $z=1.99$, making it one of the most distant clusters discovered (\citealt{2013Gobat}). The mass of this cluster (derived from the X-ray luminosity - mass correlation) was estimated to be $\sim$5$\times$10$^{13}$M$_{\odot}$ making it a relatively low mass cluster, and a typical progenitor to clusters seen today (\citealt{2011Gobat}).

Several studies have already been conducted into CLJ1449 showing that there is a population of passive galaxies within the cluster, which is the beginning of a red sequence of galaxies (\citealt{2013Strazullo,2016Strazullo}). Using data from the Atacama Large Millimetre Array (ALMA) and the Jansky Very Large Array (JVLA) both \cite{2018Coogan} (hereafter \citetalias{2018Coogan}) and \cite{2018Strazzullo} (hereafter \citetalias{2018Strazzullo}) have found that the very centre of this cluster is still actively forming stars. \citetalias{2018Strazzullo} showed that within the central 200 kpc region, stars are forming at a rate of 700 M$_{\odot}$yr$^{-1}$. However \citetalias{2018Coogan} argued that based on the gas depletion times this SF cannot be maintained and the gas will soon be used up on short time scales. It has also been suggested in \citetalias{2018Coogan} that the main cause of the high SF is down to mergers between cluster galaxies.

Due to the facilities used, these studies were only limited to the very central region of the cluster. In this paper we present new SCUBA-2 observations, combined with archival \textit{Herschel}, optical and NIR data. This allows us to observe both the entire cluster and the outlying field regions to fully investigate the SF/density relation in and around CLJ1449.

Section \ref{sec:Obs} discusses the reduction of the SCUBA-2 data and the ancillary data for this cluster. In Section \ref{sec:extract} we discuss how we created our source catalogue and extract fluxes from our maps,  whilst in \ref{sec:RS} we discuss determining cluster membership via redshifts. Section \ref{sec:SFR} discusses how we calculate the SFR of our galaxies and observe the SF/density relation and compare to other clusters.

In the analysis we assume a $\Lambda$CDM cosmology with H$_0$=70\,kms$^{-1}$Mpc$^{-1}$, $\Omega_{\Lambda}$=0.7 and $\Omega_m$=0.3. The scale at redshift 2 is 8.5\,kpc\,arcsec$^{-1}$. All magnitudes are given in the AB system, and all uncertainties are given at the 1\,$\sigma$ level.

\section{Observations and Reductions}\label{sec:Obs}

\subsection{SCUBA-2}\label{sec:scuba2}

The SCUBA-2 observations were carried out over 6 nights between April 2015-March 2016 as part of projects M15AI51 and M16AP047 (PI Gear). A total integration time of 8 hours was performed using 3\,arcmin daisy scans, with all the observations being carried out in good grade one weather ($\tau_{225}\leq$ 0.05).

Both the 450\,$\mu$m and 850\,$\mu$m data were reduced using the Dynamic Iterative Map Maker (\textsc{DIMM}) within the Sub-mm User Reduction Facility (\textsc{SMURF}, \citealt{2013Chaplin}, \citealt{2013Jenness}). A brief description of the reduction process is given here, for a full overview see \cite{2013Chaplin}.

The first steps involve down sampling and flat fielding of the raw data, and scaling it so the units are in pW. The \textsc{DIMM} then enters an iterative process where it assumes the map is a linear combination of: 
\begin{enumerate}
\item A common mode signal present in all bolometers. This is usually caused by sky noise, and ambient thermal emission;
\item The astronomical signal, which is attenuated by atmospheric extinction;
\item A noise term not accounted for by either (i) or (ii).
\end{enumerate}
The map maker iteratively solves for these and refine them until either a certain number of iterations have passed, or the map has converged. Convergence is determined when either a set number of iterations has passed, or the mean difference between maps falls below a certain value. What is left is a map that consists only of the astronomical signal (corrected for extinction) plus noise.

To account for atmospheric fluctuations, the data has to be filtered within the frequency domain. Since we are not concerned about extended structure a harsh filter can be applied. We filter out frequencies that represent scales larger than 150\,arcsec. Throughout the reduction process any bolometers that deviate massively from the average are flagged as bad and do not contribute to the final map. To help identify any faint sources we ran the match filter recipe on our maps using the \textsc{PICARD} package. This first convolves the map with a large Gaussian and then subtracts it to help remove any remaining large scale structure. This map is then convolved with the beam to help with the detection of any point sources that match the size of the beam (8\,arcsec at 450\,$\mu$m and 15\,arcsec at 850\,$\mu$m).
 
\begin{figure*}
\centering
\includegraphics[scale=0.7]{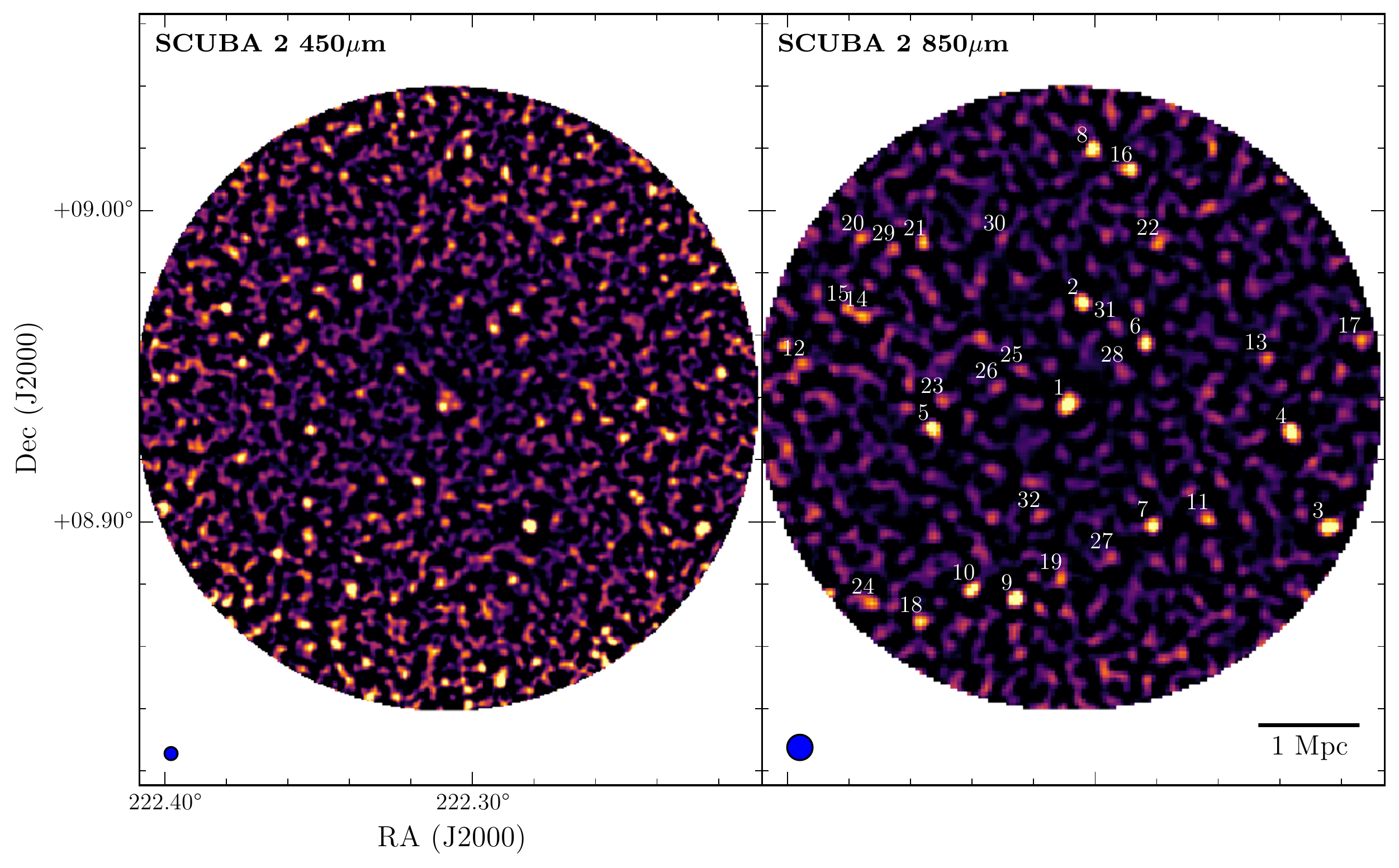}
\caption{The SCUBA-2 maps of CLJ1449 with a radius of 3\,Mpc. The FWHM size of the PSF can be found in the lower left corner. The positions of all 32 SCUBA-2 sources with a S/N greater than 4 are labeled on the 850\,$\mu$m image, with the numbering being the same as that in Table \ref{table:850_sources} }\label{fig:Imagecomp}
\end{figure*}

Calibration was done by applying extinction models outlined in \cite{2013Dempsey} to sources of a known brightness. This in turn generates a Flux Conversion Factor (FCF) which converts the raw map units of pW to the usable mJy/beam. There are standard FCF values derived from hundreds of observations, however because the 450\,$\mu$m is very sensitive to changes in the atmosphere, changes in atmospheric conditions could cause a deviation from the standard FCF. It was therefore decided to calibrate each observation individually based on each nights calibrators. Our average FCFs were 552\,Jy beam$^{-1}$pW$^{-1}$ for 850\,$\mu$m and 460\,Jy beam$^{-1}$pW$^{-1}$ for 450\,$\mu$m. Once all the observations were calibrated, they were mosaicked, and the resulting image cut to a radius of 360\,arcsec to remove the noisy edges. The final rms values were 0.96\,mJy beam$^{-1}$ for 850\,$\mu$m and 4.27\,mJy beam$^{-1}$ for 450\,$\mu$m. We compare our calibration to that of using the standard FCF values and found the noise is larger, being 1.12\,mJy beam$^{-1}$ at 850\,$\mu$m and 5.16\,mJy beam$^{-1}$ at 450\,$\mu$m. All FCF values contain an additional 10\% to account for flux lost during the reduction procedure. This was calculated by inserting fake sources of known brightness into the raw time series and comparing the flux before and after the reduction. The images can be seen in Figure \ref{fig:Imagecomp}.

\subsection{Herschel}\label{sec:Herschel}
To complement the SCUBA-2 data we have used publicly available data from both the Photoconductor Array Camera and Spectrometer (PACS, \citealt{2010Poglitsch}) and  Spectral and Photometric Imaging Receive (SPIRE, \citealt{2010Griffin}).
The PACS observations were acquired in July 2011 (ObsId 1342224474,1342224475, PI Gobat). Both 100\,$\mu$m and 160\,$\mu$m were observed over 10 hours in large scan map mode. The noise for these maps are 1.9 and 3.2\,mJy and achieve a resolution of 7.7\,arcsec and 12\,arcsec for 100 and 160\,$\mu$m respectively. All 3 SPIRE bands were observed with an integration time of 4 hours in January 2013 (ObsId 1342259448, PI Dannerbauer). The noise in the maps were 4.7, 5.5 and 6.2\,mJy with a resolution of 18\,arcsec, 26\,arcsec and 36\,arcsec for at 250, 350 and 500\,$\mu$m respectively. Both the PACS and SPIRE level 2.5 data products were downloaded from the \textit{Herschel} ESA archive\footnote{http://archives.esac.esa.int/hsa/}.

\subsection{ALMA}\label{sec:alma}

As part of cycle 1 the central area of the cluster was observed at 870\,$\mu$m using ALMA (Project code 2012.1.00885.S, P.I Strazullo). The cluster was observed for 2.5 hours and only covers a small area (0.3\,arcmin$^{2}$). The map reaches a noise of 70\,$\mu$Jy beam$^{-1}$ and has a resolution of $\sim$1\,arcsec. For more information see \citetalias{2018Coogan}. The data was acquired from the ALMA ESO science archive\footnote{http://almascience.eso.org/aq/}.

Even though the data only covers the central region of the cluster (see Figure \ref{fig:FOV}), the higher resolution means that we will be able to detect confused source members that cannot be resolved in the SCUBA-2 images. This will be discussed further in Section \ref{sec:extract}.

\subsection{JVLA - S Band}
The S band radio data consist of observations from 2012 of JVLA S band (2-4GHz, project code:
12A-188, PI: V.Strazzullo.) using 16 spectral widows (128MHz bandwidth, 64 channels). The raw data was downloaded from the NRAO's VLA archive\footnote{https://science.nrao.edu/facilities/vla/archive/}, and each observation separately reduced using NRAO's Astronomical Image Processing System (AIPS), following the standard VLA data calibration procedure described in the AIPS cookbook. For each observation, the calibrated UV data of the target field have been thoroughly flagged for radio-frequency interference and separated without averaging the frequency channels to avoid bandwidth smearing (chromatic aberrations). All of the resulting UV data sets have been combined for the final imaging, which consisted of a single phase-only and a final amplitude-and-phase calibration.

\subsection{Spitzer}
Complimentary IR data of CLJ1449 was obtained with \textit{Spitzer's} Infrared Array Camera (IRAC, \citealt{2004Fazio}). All 4 bands (3.6, 4.5, 5.8, and 8\,$\mu$m, PI Giovanni, ObsId 4393984) were observed and used. Each of the IRAC bands cover the same area (27 arcmin $\times$ 22 arcmin) but due to offset between arrays, only the 3.6\,$\mu$m and 5.8\,$\mu$m data cover the same area as that in the SCUBA-2 maps. As part of the \textit{Spitzer} warm mission the cluster was re-observed at 3.6\,$\mu$m and 4.5\,$\mu$m (PI Gobat, ObsId 42576640) and when combined with the pre-existing data complete coverage of the cluster is achieved for 3 of the 4 bands.

\subsection{Additional Data}
\begin{figure*}
\centering
\includegraphics[scale=0.8]{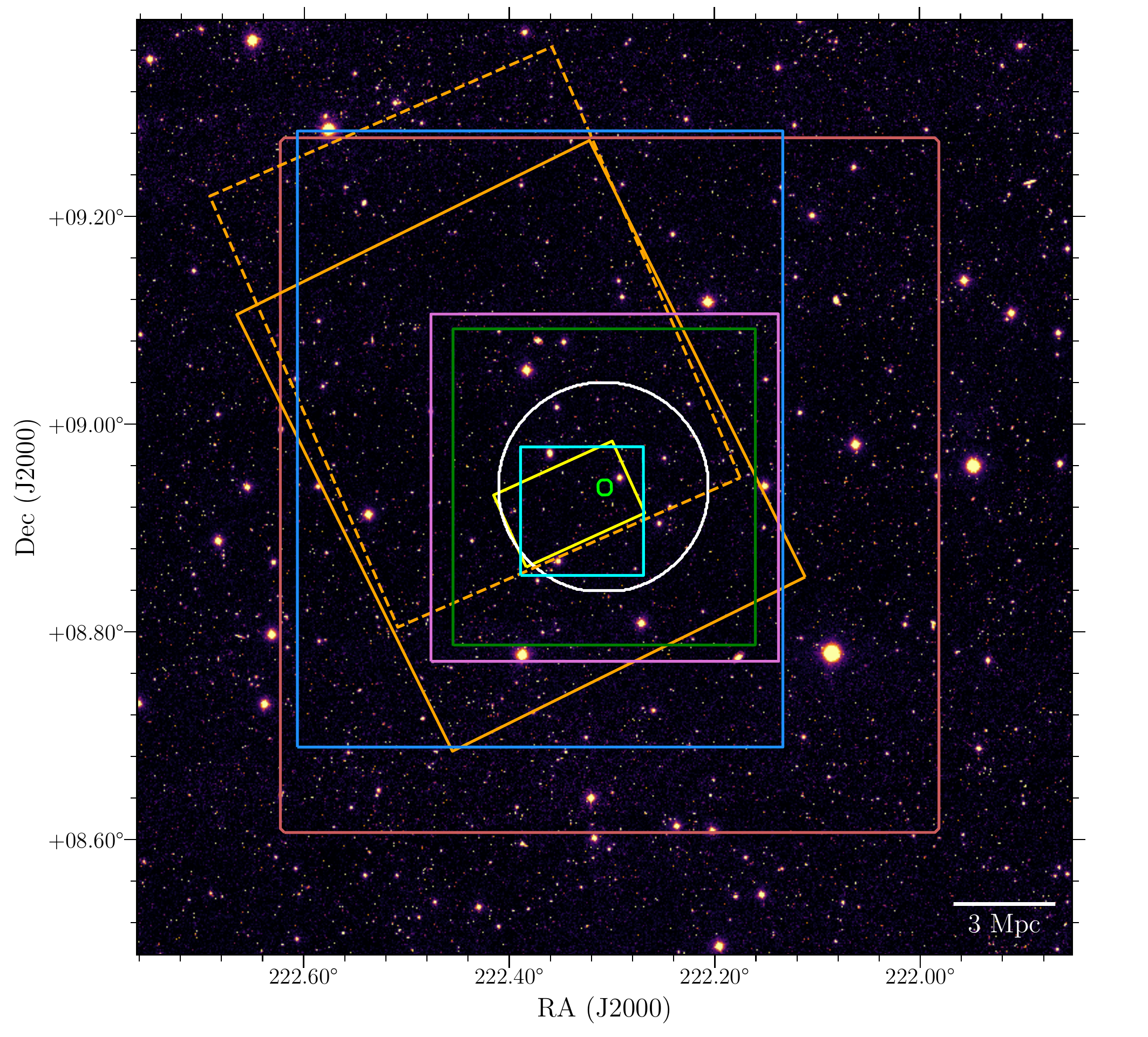}
\caption{FOVs of all the data used in this analysis. The white circle represents the SCUBA-2 coverage (450/850\,$\mu$m), whilst the small lime green circle represents the ALMA coverage. The yellow rectangle is the MOIRCS and ISAAC coverage (Y, H, J, and Ks bands), whilst the cyan square is the FORS 2 coverage (U and V band). The green square shows the PACS coverage (100/160\,$\mu$m) and the purple square shows the radio data (10\,cm). The solid orange rectangle shows the IRAC coverage at 3.6,4.6 and 5.8\,$\mu$m, with the dashed orange rectangle showing the 8\,$\mu$m coverage. The blue square shows the Suprime cam coverage (B, I and z bands) and the red square shows the SPIRE coverage (250/350/500\,$\mu$m). The background is a g band image from SDSS.}\label{fig:FOV}
\end{figure*}

We also utilise archival optical and NIR maps for CLJ1449. Deep $B$, $I$ and $z$ were taken with Suprime-cam (\citealt{2002Miyazaki}) on the Subaru telescope achieving 5\,$\sigma$ magnitudes depths of 26.95, 26.03 and 25.91 respectively. We also have NIR data ($Y$, $H$, $J$ and $Ks$) taken with both the Multi-Object Infrared Camera and Spectrograph (MOIRCS, \citealt{2006Ichikawa}, \citealt{2008Suzuki}) on Subaru and the Infrared Spectrometer And Array Camera (ISAAC, \citealt{1998Moorwood}) on the Very Large Telescope (VLT). These images reach depths at 5\,$\sigma$ of 25.64, 25.47, 23.66 and 24.74 for $Y$, $H$, $J$ and $Ks$ respectively. Finally we have $U$ and $V$ band data taken with the FOcal Reducer and low dispersion Spectrograph (FORS, \citealt{1998Appenzeller}) also on the VLT, reaching 5\,$\sigma$ depths of 28.1 and 26.52 for $U$ and $V$. For more information on these data sets we refer to \cite{2006Kong}, \cite{2011Gobat} and \cite{2013Strazullo}. A summary of all data used in this analysis can be found in Table \ref{table:summary}, and the fields of view (FOV) for all the maps used can be found in Figure \ref{fig:FOV}.

\begin{table}
\caption{Summary of the data-sets used in our analysis.}  
\label{table:summary}     
\small
\centering           
\begin{tabular}{cccc} 
\hline\hline   
Telescope          & Instrument    &        Observed Band & FOV\\
\hline                         
VLT               &  FORS 2    &    U, V & 7 arcmin $\times$ 7 arcmin\\
Subaru                   &   Suprime    &  B,I,z & 27 arcmin $\times$ 35 arcmin \\
Subaru                   &   MOIRCS    &  Y,H & 7 arcmin $\times$ 4 arcmin\\
VLT              &    ISAAC    &  J, Ks & 7 arcmin $\times$ 4 arcmin\\
Spitzer                 &   IRAC    &  3.6/4.5/5.8/8\,$\mu$m & 27 arcmin $\times$ 22 arcmin \\
Herschel              &   PACS    &  100/160\,$\mu$m & 15 arcmin $\times$ 15 arcmin\\
Herschel              &   SPIRE    &  250/350/500\,$\mu$m & 35 arcmin $\times$ 35 arcmin\\
JCMT              &   SCUBA-2    &  450/850\,$\mu$m & 113\,arcmin$^{2}$\\
ALMA & - & 870\,$\mu$m & 0.3\,arcmin$^{2}$\\
JVLA & - & 10\,cm & 20 arcmin $\times$ 20 arcmin\\
\hline
\end{tabular}
\end{table}

\begin{figure*}
\centering
\includegraphics[scale=0.8]{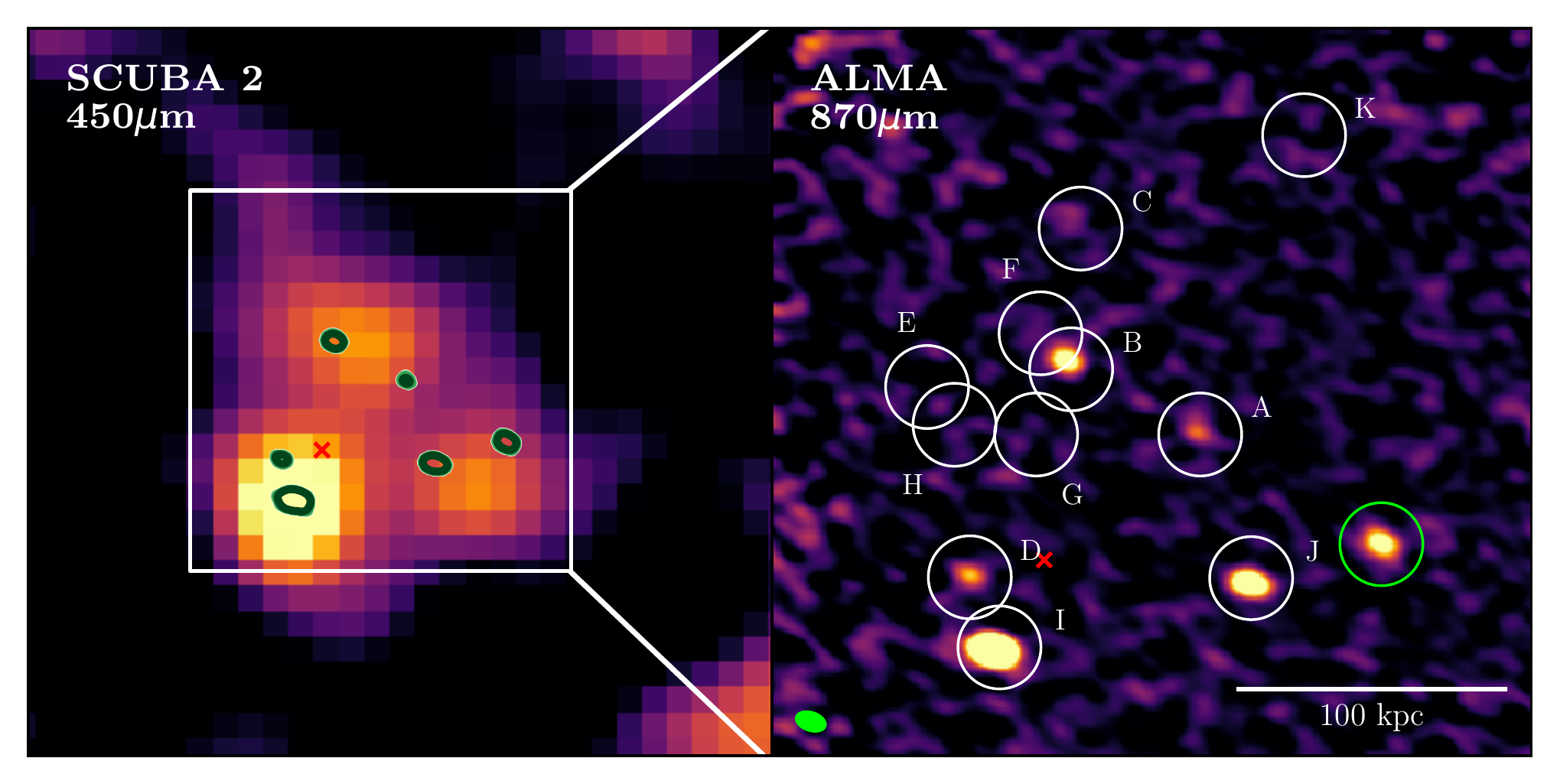}
\caption{The ALMA 870\,$\mu$m continuum map of the core of the cluster. The ALMA data resolves the SCUBA-2 450\,$\mu$m map into 6 individual sources rather then 3. The size of the PSF can be see in the bottom left corner. The letter represents the sources in Table \ref{table:submm}, and are based on the designations from \citetalias{2018Coogan}. A low redshift object is known to contaminate the cluster, and is illustrated by the green circle. The red cross represents the centre of the x-ray emission from \citealt{2011Gobat}.}\label{fig:ALMA}
\end{figure*}

\section{Source Identification}\label{sec:extract}

To identify any potential sub-mm sources we search for any sources in the SCUBA-2 850\,$\mu$m map that have a signal to noise ratio (S/N) greater than 4. This uncertainty level was determined by inserting fake sources into a jackknife map, made by inverting half of the observations and co-adding. This produces a map with the same noise properties as the final map. Fake sources were then randomly inserted into the map (10$^{5}$ sources inserted in batches of ten, with fluxes picked from a uniform distribution) and \textsc{Sextractor} (\citealt{1996Bertin}) was then used to extract these sources.

It was found that with a S/N of 3.5 we are over 80\% complete at $\sim$5\,mJy. However the number of fake sources (i.e. sources caused entirely by noise) was very high, with there being 8 sources detected. At a S/N of 4 there was only 1 fake source, which is expected in a map of this size with Gaussian fluctuations. Therefore a S/N cut of 4 was selected.

In the SCUBA-2 850\,$\mu$m map we detected 32 sources that have a S/N greater than 4, with the positions of all these sources given in Table \ref{table:850_sources}, and their locations illustrated in Figure \ref{fig:Imagecomp}. When we looked at the 450\,$\mu$m map we found that there were 18 sources with a S/N greater than 4. When cross matching these sources with the 850\,$\mu$m ones we found only 11 sources matched. This low matching rate was also found by the SCUBA-2 Cosmology Legacy Survey (S2CLS). \cite{2013Casey} found that in the COSMOS field there was only a $\sim$30\% matching rate when comparing 450 and 850 sources at the same S/N cut. This implies that the 450\,$\mu$m population is intrinsically different than that of the 850\,$\mu$m and indicates that different redshift or luminosity populations are being probed.

From the number counts of the S2CLS, in a field of the same size, at this sensitivity we would expect to see 10$\pm$3 sources at 850\,$\mu$m (\citealt{2017Geach}). This means that our cluster field is more than three times over-dense.

\begin{table*}
\caption{Catalogue of sources from the SCUBA-2 850\,$\mu$m map that have a S/N greater than 4.}  
\label{table:850_sources}     
\small
\centering           
\begin{tabular}{ccccccccccc} 
\hline\hline   
ID         & RA    &  DEC & $r_c$ & S/N$_{850}$ & S/N$_{450}$ \\
&&& (Mpc) && \\
\hline
850\textunderscore1 &   222.30882& 8.93770&0.0 & 10.6& 7.7\\
850\textunderscore2 &   222.30423& 8.97015&0.958 &9.8& 4.0 \\
850\textunderscore3 &   222.22436& 8.89832 &2.864 &8.8& 6.0\\
850\textunderscore4 &   222.23642 & 8.92864&2.229 &8.1& 3.5 \\
850\textunderscore5 &   222.35272&  8.92960&1.392 &8.0& 6.6\\
850\textunderscore6 &   222.28385 & 8.95749&0.938 &7.8 & 3.4\\
850\textunderscore7 &   222.28147&  8.89870&1.489 &7.4& 10.8\\
850\textunderscore8 &   222.30102&  9.01891&2.456 &6.6& 3.5\\
850\textunderscore9 &   222.32561&  8.87515&2.033 &6.5& 3.1\\
850\textunderscore10 &  222.33980&  8.87835&2.100 &5.7& 5.6\\
850\textunderscore11 &  222.26360& 8.90100&1.803 &5.3& <2\\
850\textunderscore12 &  222.39497& 8.95028&2.679 &5.0& <2\\
850\textunderscore13 &  222.24476& 8.95221&1.990 &4.9& 2.5\\
850\textunderscore14 &  222.37475& 8.96610&2.198 &4.9& <2\\
850\textunderscore15 &  222.38069& 8.96798&2.389 &4.8& 6.8\\
850\textunderscore16 &  222.28853& 9.01256&2.331 &4.8& 4.4\\
850\textunderscore17 &  222.21441& 8.95775&2.937 &4.8& 2.6\\
850\textunderscore18 &  222.35692& 8.86788&2.646 &4.8& 3.6\\
850\textunderscore19 &  222.31144& 8.88167&1.766 &4.7& 3.0\\
850\textunderscore20 &  222.37582& 8.99054&2.602 &4.5& 2.9\\
850\textunderscore21 &  222.35558& 8.98888&2.104 &4.4& 5.0\\
850\textunderscore22 &  222.27984& 8.98906&1.762 &4.3& <2\\
850\textunderscore23 &  222.34994& 8.93833&1.276 &4.3& 2.8\\
850\textunderscore24 &  222.37243& 8.87388&2.806 &4.3& 2.9\\
850\textunderscore25 &  222.32408& 8.94833&0.558 &4.3& 5.1\\
850\textunderscore26 &  222.33233& 8.94313&0.746 &4.2 & 2.2\\
850\textunderscore27 &  222.29483& 8.88833&1.613 &4.2& 3.3\\
850\textunderscore28 &  222.29146& 8.94833&0.589 &4.1& 2.8\\
850\textunderscore29 &  222.36570& 8.98721&2.295 &4.1& <2\\
850\textunderscore30 &  222.32970& 8.99055&1.707 &4.1& <2\\
850\textunderscore31 &  222.29371& 8.96277&0.851 &4.0& 6.0\\
850\textunderscore32 &  222.31845& 8.90166&1.191 &4.0& 2.6\\

\end{tabular}
\end{table*}

\subsection{Source Confusion}

Due to the large beam size of our SCUBA-2 data, assigning optical and NIR counterparts is difficult as several galaxies could be residing in one beam. An example of this can be seen with the SCUBA-2 images in Figure \ref{fig:Imagecomp}. In the 850\,$\mu$m map (beam size 15\,arcsec) the central region is seen as one source, but the 450\,$\mu$m (beam size 8\,arcsec) shows this one source resolved into 3 separate sources. The severity of this issue is fully realised when we compare the SCUBA-2 data to the ALMA 870\,$\mu$m data, with a resolution of $\sim$1\,arcsec. Figure \ref{fig:ALMA} shows that the 3 sources in the 450\,$\mu$m data actually is 6 individual sources, and this is only seen due to the high resolution of ALMA. It should be noted that \citetalias{2018Coogan} actually identify 11 galaxies within the central 200\,kpc based on the positions of CO(4-3) detections.

Due to the large beam sizes of both SCUBA-2 and \textit{Hershcel} there is a high probability that several sources may be blended into one individual source. This makes associating the FIR/sub-mm sources to optical and NIR sources complicated, as we do not know which optical and NIR sources are associated with the FIR/sub-mm ones.

\subsection{De-Blending Sub-mm Images}\label{sec:XID+}

To de-blend our images and get accurate flux measurements we used the Bayesian inference tool XID+\footnote{https://github.com/H-E-L-P/XID\textunderscore plus} (\citealt{2017Hurley}). XID+ uses the positional data from a tracer of dusty star formation and explores the full posterior function to extract fluxes from confused maps. To assure the most accurate results from XID+, high resolution positional data is required from a wavelength that traces dust obscured star formation. For the central region we use the ALMA map and the 11 sources detected at 870\,$\mu$m by \citetalias{2018Coogan}. We also included the lower right source (circled green in Figure \ref{fig:ALMA}) even though this is a known low redshift object. This is because we need it for the de-blending process, but it is excluded from the rest of the analysis.

For the rest of the region, we had to use a different tracer. Good examples are either 24\,$\mu$m (e.g. \citealt{2009Marsden}, \citealt{2009Pascale}, \citealt{2010Elbaz}, \citealt{2012Bethermin}) or radio (e.g. \citealt{2010Ivison}, \citealt{2010Magnelli}, \citealt{2015Basu}, \citealt{2016Rujopakarn}, \citealt{2017Delhaize}). Even though a MIPS 24\,$\mu$m map was available, the beam size is still large ($\sim$6\,arcsec) and galaxies will still be blended. Instead we used the 10\,cm (3\,GHz) JVLA map as our tracer, as it has resolution comparable to that of our ALMA map. 

We identified 319 significant sources in the radio map (within the same area of the SCUBA-2 maps), and combine these with the 12 ALMA sources. These sources were then passed through XID+ to obtain fluxes for both the PACS and SPIRE maps. A specially modified version of XID+ was used to work with the SCUBA-2 maps, so we also have de-blended fluxes for both 450\,$\mu$m and 850\,$\mu$m.

\subsection{Association With Radio Sources }

We then went about associating each of our SCUBA-2 sources with at least one radio source (and therefore can associate with optical and NIR sources). To determine the most likely radio source we use the standard Poissonian probability of positional match similar to the method outlined in \cite{1986Downes}. This value $P$ gives us the probability of a source \textit{not} being associated with the SCUBA-2 source and is given by
\begin{equation}
P=1-\exp{(-\pi\,n\,\theta^2)},
\end{equation}
where $n$ is the number density of sources in the radio map, and $\theta$ is the separation between the SCUBA-2 source and the radio source. We found all radio sources within 15\,arcsec of the SCUBA-2 source, and calculated the $P$ value for all of them. For galaxies that had similar $P$ values we picked the source that was reddest in the IRAC bands, as it has been shown that sub-mm galaxies tend to be redder at these wavelengths (e.g. \citealt{2006Ashby}, \citealt{2008Yun}, \citealt{2009Hainline}). We did not worry about the central source (source 850\textunderscore1) as we use the ALMA sources from \citetalias{2018Coogan}, which have been shown to have association with optical and NIR sources.

Using the prescriptions laid out in \cite{2002Ivison}, \cite{2009Chapin} and \cite{2016Chen}, we considered a reliable match to have a $P$\,$\leq$\,0.05 and a tentative match to have 0.05$\leq P\leq$0.1. Anything with a $P>$0.1 is considered an untrustworthy match and removed. We found that all but 2 sources had either reliable or tentative matches, with 850\textunderscore12 and 850\textunderscore19 having sources too far to be considered significant. Even though 850\textunderscore18 and 850\textunderscore28 had low $P$ values, in the optical and NIR images they were obscured by nearby stars and hence removed from the sample. Finally 850\textunderscore30 was removed because there was no radio source within 15\,arcsec. Applying these cuts we ended up with 37 sources that we can calculate redshifts for, and determine if they are likely to be within the cluster or not.

\section{Redshift Determination}\label{sec:RS}

To assign cluster membership we estimated redshifts for our remaining 37 galaxies, and those that had a redshift consistent with the cluster were considered members. However due to a lack of spectral data beyond the core region, we have to rely on photometric methods to determine the redshift for our sources. We performed aperture photometry on the optical and NIR maps (based on the radio positions), and accounted for difference in resolutions by applying aperture corrections. The maps were also calibrated by using stars of known brightness. This resulted in a 13 band catalogue spanning from U band ($\sim$\,0.3\,$\mu$m) to 8\,$\mu$m. However due to differing map sizes, only 6 bands (B, z, I, 3.6, 4.5 and 5.8\,$\mu$m) had the same coverage as the SCUBA-2 maps, meaning our catalogue is incomplete (especially at NIR wavelengths).

\subsection{Photometric Redshifts Using Optical and NIR Data}

\begin{figure}
\centering
\includegraphics[scale=0.45]{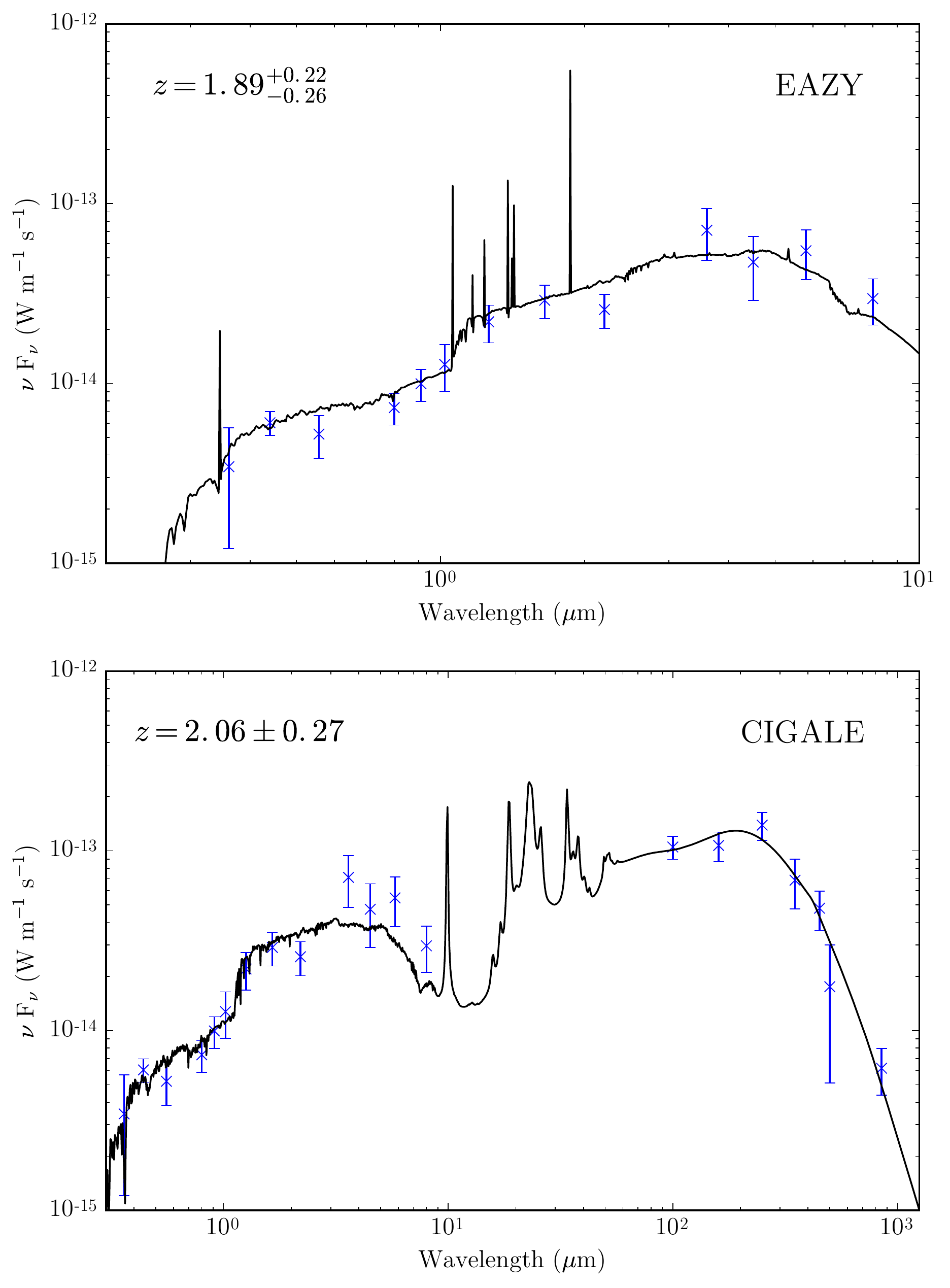}
\caption{Both the \textsc{EAZY} SED  (top) and \textsc{CIGALE} SED (bottom) for source 850\textunderscore25 }\label{fig:SED}
\end{figure}

\begin{figure}
\centering
\includegraphics[scale=0.7]{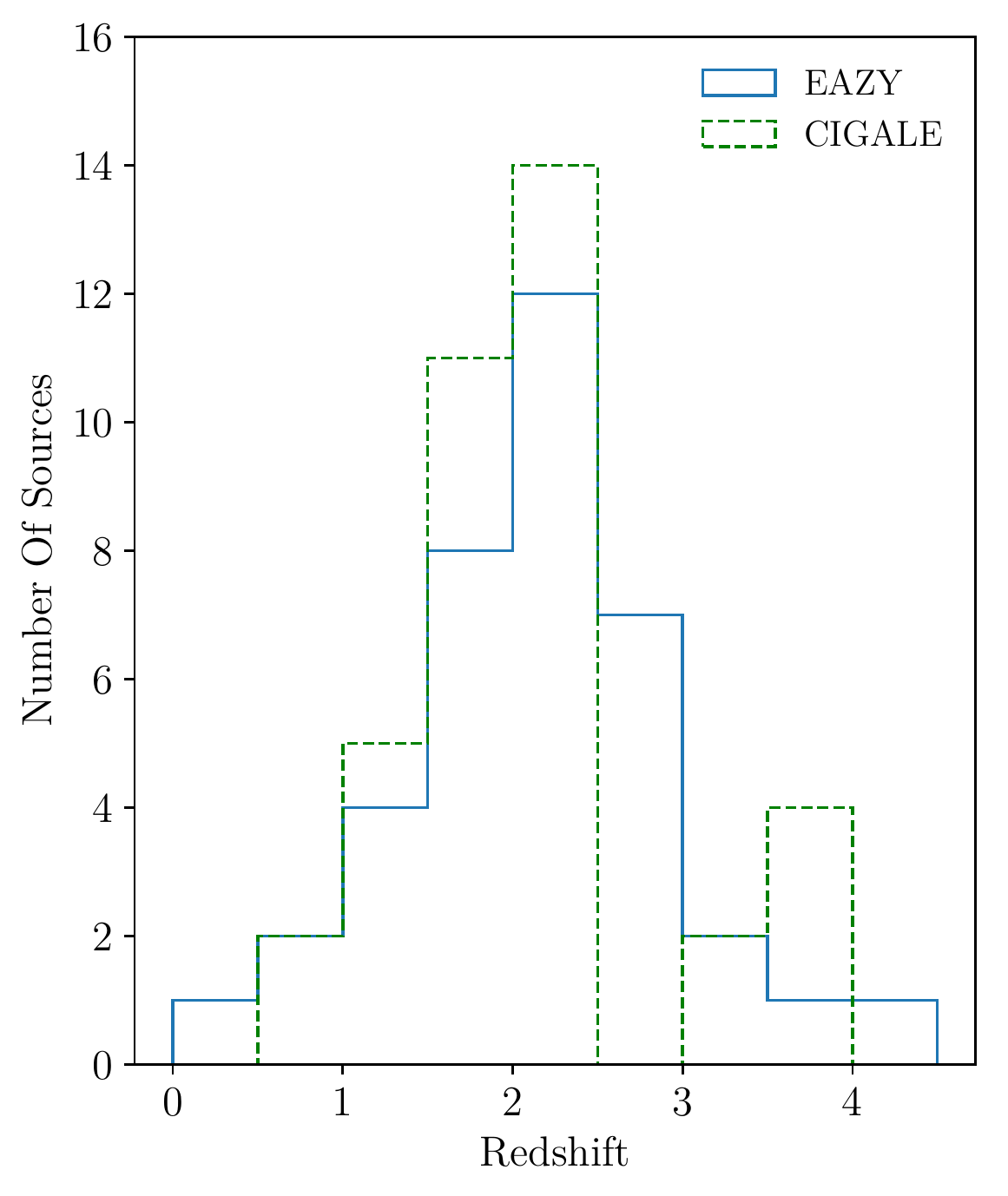}
\caption{The photometric redshift distributions for both \textsc{EAZY} and \textsc{CIGALE}.}\label{fig:z_hist}
\end{figure}

To calculate our photometric redshifts we first used the template fitting code \textsc{EAZY}\footnote{https://github.com/gbrammer/eazy-photoz/} using the standard set of templates  (\citealt{2008Brammer,2011Brammer}, \citealt{2011Whitaker}). We can see an example of the fit from \textsc{EAZY} in Figure \ref{fig:SED}, whilst Figure \ref{fig:z_hist} shows the distribution of redshifts calculated with \textsc{EAZY}.

The main issue when using \textsc{EAZY} is our incomplete catalogue. Whilst \textsc{EAZY} gives accurate results in areas with complete data, such as in the cluster centre (with $\Delta$z being on average\,$\sim$\,0.2), in regions where there is no NIR data and just the Subaru and \textit{Spitzer} data, $\Delta$z on average is\,$\sim$\,1 removing any precision needed in identifying cluster members. Using the optical and NIR data on its own is therefore not the most ideal situation, so we also looked at methods that included FIR/sub-mm data.

\subsection{Photometric Redshifts Including Sub-mm Data}

Using FIR/sub-mm data to estimate photometric redshifts has been studied several times  (e.g. \citealt{2013Pearson}, \citealt{2016Ivison}, \citealt{2018Bakx}) and is ideal for situations where there is little or no optical and NIR counterparts to the sub-mm sources. However, these results can have significant uncertainties ($\pm$0.5 in some cases), and on their own would not have the accuracy to place them within the cluster. This is especially true with XID+, as areas where several sources exist (such as the centre) the errors on the fluxes are significant (up to 50\% in some cases).

To calculate photometric redshifts using both optical, NIR, FIR and sub-mm data we use The Code Investigating GALaxy Emission (\textsc{CIGALE}\footnote{https://cigale.lam.fr/}, \citealt{2018Boquien}). \textsc{CIGALE} is a dust energy balancing code, which balances any energy lost via dust attenuation to that of the emission caused by the dust. This approach is more advantageous as even with a small/incomplete optical data set, reasonably accurate results can be estimated as long as the FIR/sub-mm data is complete. The main advantage of \textsc{CIGALE} over similar codes is the ability to leave redshift as a free parameter, and compute photometric redshifts from the full wavelength range.

Another advantage of \textsc{CIGALE} is the flexibility in models that can be selected. With \textsc{CIGALE} the user can determine the models and parameters used (e.g. what dust emission model is used, what star formation history, etc) which allows for greater variability. We decided to use the same parameters used in the Herschel Extragalactic Legacy Project (HELP, \citealt{2016Vaccari}). These parameters were selected as they cover a wide range of models and suitable for most galaxy types, and include a delayed star formation history (with additional burst), single stellar population models from \cite{2003Bruzaul}, dust attenuation from \cite{2000Charlot} and a \cite{2007Draine} dust emission model. For more information see \cite{2018Malek}.

\textsc{CIGALE} was run for our 37 galaxies using the HELP settings, except for redshift which was kept as a free parameter. With the added data we found that the uncertainties were smaller, especially for those galaxies that had significant uncertainties with \textsc{EAZY}. Again an example of a fit generated from \textsc{CIGALE} can be seen in Figure \ref{fig:SED} and distribution of redshifts can be seen in Figure \ref{fig:z_hist}. Comparisons between all three redshifts measures can be seen in Figure \ref{fig:Scatter}.

\subsection{Cluster Membership}

\begin{figure}
\centering
\includegraphics[scale=0.85]{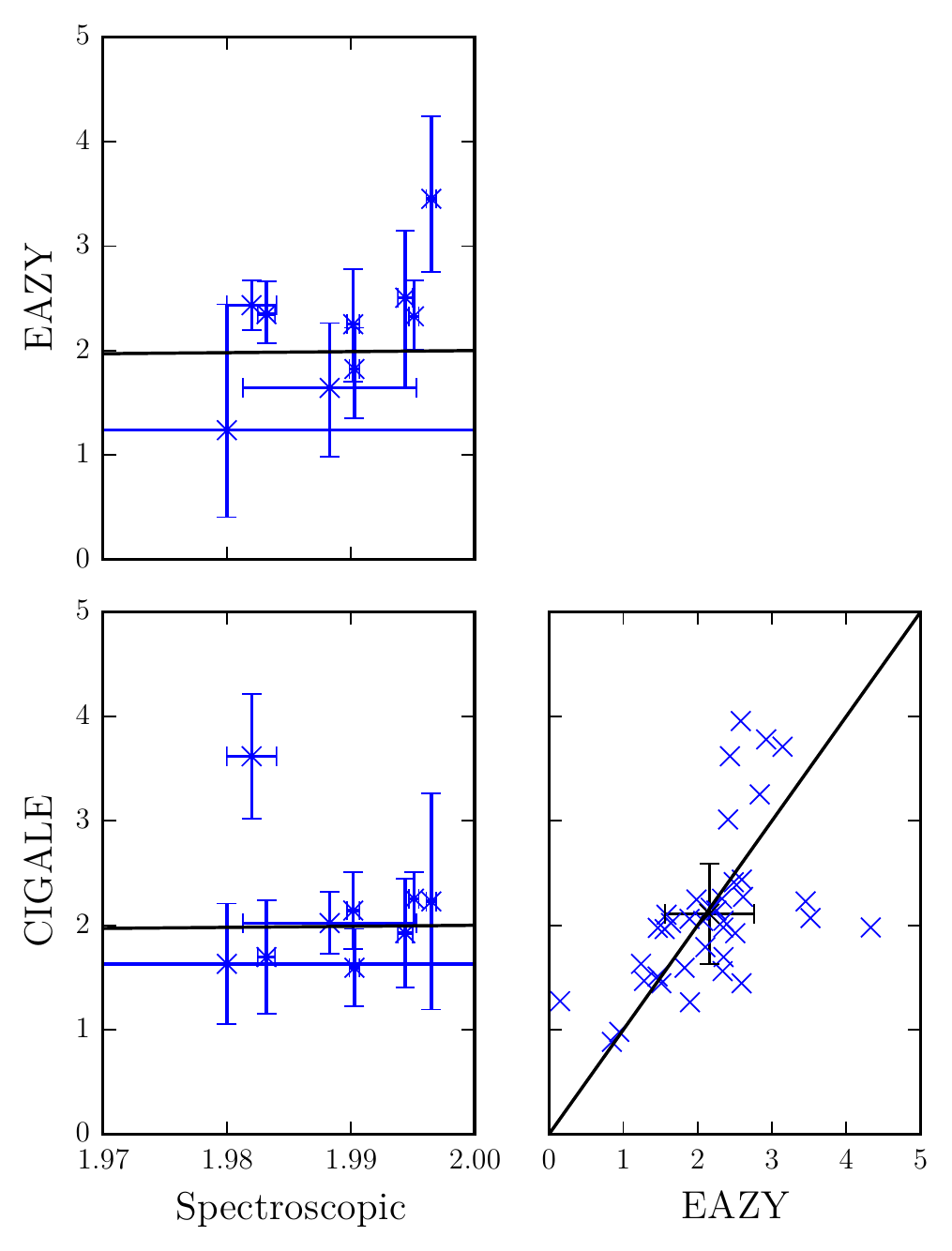}
\caption{Comparing all three redshift determination methods. The black point shows the median value for both \textsc{CIGALE}, \textsc{EAZY} and errors.}\label{fig:Scatter}
\end{figure}

To determine the cluster membership we compared the photometric values to the values of those sources with spectral redshifts (Figure \ref{fig:Scatter}). The only sources that had spectral redshifts are nine of the central ALMA sources discussed in \citetalias{2018Coogan}. To determine a range of redshifts that could indicate membership, we compared the scatter between the photometric and spectroscopic redshifts. We found that for both \textsc{CIGALE} and \textsc{EAZY} the scatter was $\sim 0.2$. Therefore it was decided that all galaxies within 2$\sigma$ of this scatter was considered a possible cluster member (i.e. any galaxy with a redshift between 1.6 and 2.4 is considered to be potentially in the cluster).

We use a combination of both the \textsc{EAZY} and \textsc{CIGALE} redshifts to help determine cluster memberships. Based on these redshifts we decided on 3 categories of cluster membership, with the first having the highest probability of being in the cluster. These galaxies either have a spectroscopic redshift or meet our redshift cut in both the \textsc{EAZY} \textit{and} \textsc{CIGALE} redshifts. The second category are galaxies that have a much less chance of actually being a cluster member. These are galaxies that have a redshift matching our cut in either \textsc{EAZY} \textit{or} \textsc{CIGALE}. The final category is galaxies which are very likely to not be in the cluster, and do not have a redshift in  \textsc{EAZY} or \textsc{CIGALE}. These galaxies are excluded from the rest of our analysis.

\begin{figure*}
\centering
\includegraphics[scale=0.6]{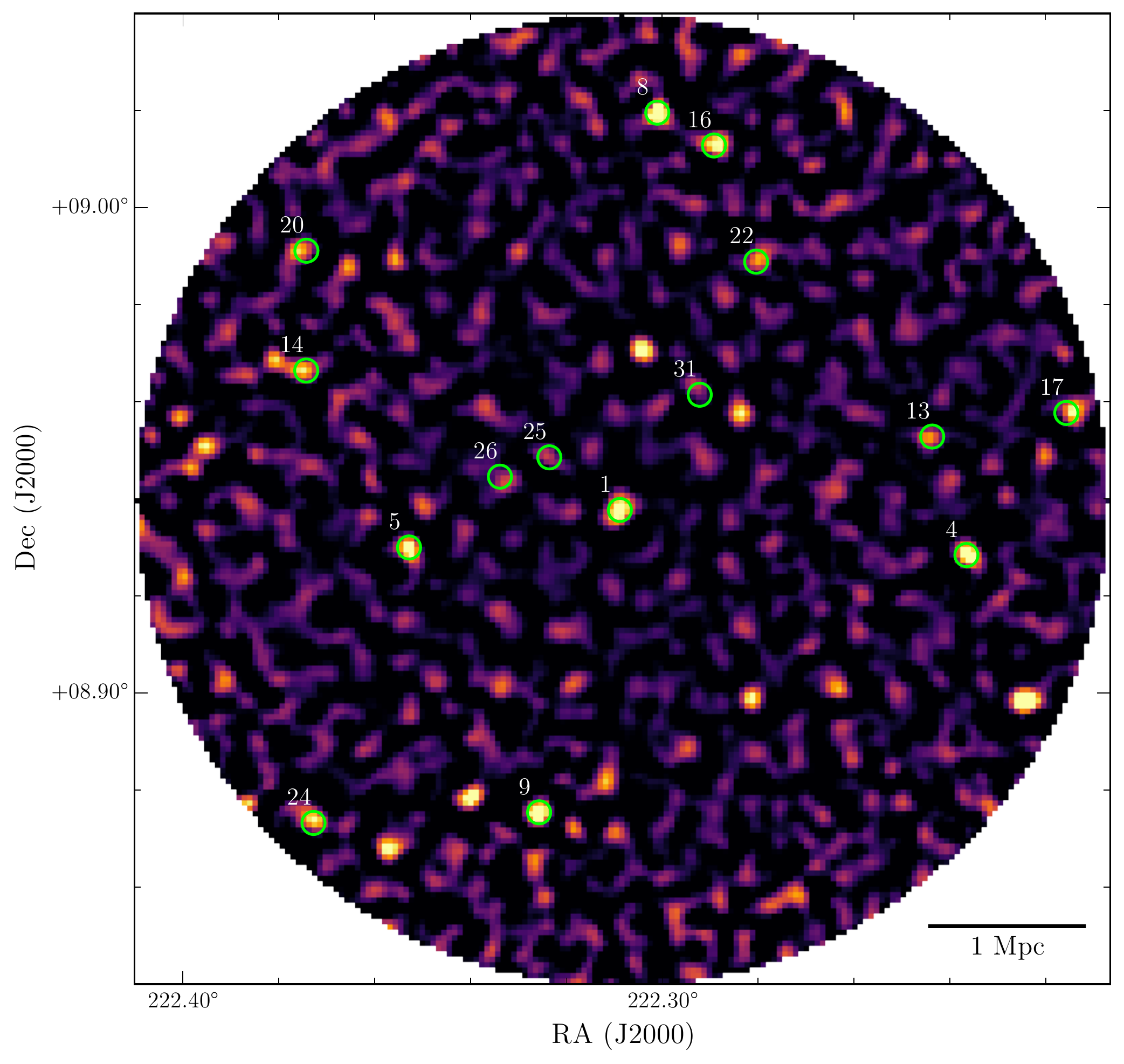}
\caption{Location of the final galaxies identified as possibly being within the cluster. It should be noted 850\textunderscore1 actually consists of 10 members.}\label{fig:Final_Pos}
\end{figure*}

Applying this criteria we found that 16 galaxies have a high probability of being in the cluster, with nine of these having spectra and seven having redshifts confirmed by both \textsc{EAZY} \textit{and} \textsc{CIGALE}. Eight galaxies have a tentative membership, with two only being confirmed by \textsc{EAZY}, and six only being confirmed by \textsc{CIGALE}. Overall we are left with 24 galaxies that could be cluster members with their fluxes being presented in Table \ref{table:submm}, and the locations of them can be seen in Figure \ref{fig:Final_Pos}. A table of all non-cluster members can be found in Table \ref{table:NCRS}.

We note that the number of galaxies that we exclude (14 galaxies) matches up well with the number of galaxies that we would expect to find in the field (Section \ref{sec:extract}, 10$\pm$3). This gives a greater confidence that the remaining galaxies are actually associated with the cluster.

\begin{table*}
\caption{FIR/sub-mm properties of our cluster galaxies.}  
\label{table:submm}     
\small
\centering  
\scalebox{0.85}{         
\begin{tabular}{cccccccccccc} 
\hline\hline   
ID         & RA    &    DEC  & $r_c$&  $f_{100}$ & $f_{160}$ & $f_{250}$ & $f_{350}$ & $f_{450}$& $f_{500}$ & $f_{850}$ & Source\\
&&&(Mpc)&(mJy)&(mJy)&(mJy)&(mJy)&(mJy)&(mJy)&(mJy) &\\\hline                         
850\textunderscore1\textunderscore A & 222.307 & 8.9395 & 0.04 &  0.26 $\pm$ 0.26 &  0.74 $\pm$ 0.73 & 10.26 $\pm$  5.26 &  6.36 $\pm$  5.99 &  1.12 $\pm$ 1.19 &  3.63 $\pm$ 3.76 & 0.81 $\pm$ 0.6 & Spec \\
850\textunderscore1\textunderscore B & 222.309 & 8.9403 & 0.04 &  0.31 $\pm$ 0.32 &  0.77 $\pm$ 0.83 &  1.66 $\pm$  1.76 &  2.32 $\pm$  2.51 &  1.06 $\pm$ 1.17 &  1.88 $\pm$ 2.15 & 0.47 $\pm$ 0.45 & Spec \\
850\textunderscore1\textunderscore C & 222.309 & 8.942  & 0.09 &  1.35 $\pm$ 0.59 &  2.41 $\pm$ 1.46 &  2    $\pm$  1.84 &  1.22 $\pm$  1.38 &  2.19 $\pm$ 1.72 &  1.24 $\pm$ 1.4  & 0.2  $\pm$ 0.22 & Spec \\
850\textunderscore1\textunderscore D & 222.31  & 8.9378 & 0.06 &  0.43 $\pm$ 0.39 &  0.7  $\pm$ 0.83 &  6.57 $\pm$  5.26 &  6.74 $\pm$  5.88 &  0.98 $\pm$ 1.09 &  4.03 $\pm$ 4.29 & 0.31 $\pm$ 0.37 & Spec \\
850\textunderscore1\textunderscore E & 222.31  & 8.9401 & 0.07 &  0.44 $\pm$ 0.37 &  0.72 $\pm$ 0.77 &  1.53 $\pm$  1.71 &  2.01 $\pm$  2.12 &  0.58 $\pm$ 0.64 &  1.61 $\pm$ 1.95 & 0.26 $\pm$ 0.29 & Spec \\
850\textunderscore1\textunderscore F & 222.309 & 8.9407 & 0.05 &  0.69 $\pm$ 0.53 &  1.49 $\pm$ 1.32 &  1.4  $\pm$  1.64 &  1.85 $\pm$  2.08 &  1.53 $\pm$ 1.47 &  1.61 $\pm$ 1.92 & 0.37 $\pm$ 0.36 & Spec \\
850\textunderscore1\textunderscore G & 222.309 & 8.9395 & 0.03 &  0.24 $\pm$ 0.26 &  0.42 $\pm$ 0.47 &  1.61 $\pm$  1.76 &  3.48 $\pm$  3.77 &  0.58 $\pm$ 0.69 &  2.51 $\pm$ 2.88 & 0.38 $\pm$ 0.39 & Spec \\
850\textunderscore1\textunderscore H & 222.31  & 8.9396 & 0.06 &  0.41 $\pm$ 0.37 &  0.68 $\pm$ 0.68 &  1.56 $\pm$  1.71 &  1.86 $\pm$  2.21 &  0.53 $\pm$ 0.58 &  1.75 $\pm$ 1.95 & 0.25 $\pm$ 0.26 & Spec \\
850\textunderscore1\textunderscore K & 222.306 & 8.9431 & 0.14 &  0.48 $\pm$ 0.36 &  2.51 $\pm$ 1.08 &  2.3  $\pm$  1.91 &  1.16 $\pm$  1.28 &  0.99 $\pm$ 0.94 &  1.06 $\pm$ 1.16 & 0.09 $\pm$ 0.1 & Spec  \\
850\textunderscore13 & 222.244 & 8.9528 & 2.03 &  0.38 $\pm$ 0.33 &  1.85 $\pm$ 1.8  & 17.58 $\pm$  4.18 & 12.48 $\pm$  6.43 &  2.44 $\pm$ 1.87 &  6.24 $\pm$ 4.31 & 1.33 $\pm$ 0.96 & Both \\
850\textunderscore14 & 222.374 & 8.9664 & 2.19 &  3.79 $\pm$ 0.55 & 11.74 $\pm$ 1.26 & 21.31 $\pm$  2.27 & 19.66 $\pm$  2.99 &  2.04 $\pm$ 1.76 & 12.61 $\pm$ 4.17 & 4.31 $\pm$ 0.86 & Both \\
850\textunderscore17 & 222.216 & 8.9577 & 2.9  &  3.33 $\pm$ 0.55 &  4.57 $\pm$ 1.08 &  8.44 $\pm$  2.4  &  3.36 $\pm$  2.7  &  6.65 $\pm$ 3.21 &  1.42 $\pm$ 1.6  & 3.55 $\pm$ 0.92 & Both \\
850\textunderscore20 & 222.374 & 8.9911 & 2.57 &  5.73 $\pm$ 0.6  & 19.03 $\pm$ 1.43 & 36.86 $\pm$  2.86 & 32.48 $\pm$  3.75 &  5.45 $\pm$ 2.68 & 21.22 $\pm$ 4.56 & 2.57 $\pm$ 0.88 & Both \\
850\textunderscore22 & 222.28  & 8.9889 & 1.75 &  1.93 $\pm$ 0.74 &  2.39 $\pm$ 1.42 &  4.94 $\pm$  3.34 &  5.89 $\pm$  4.27 &  2.12 $\pm$ 1.8  &  4.23 $\pm$ 3.47 & 3.09 $\pm$ 1.29 & Both \\
850\textunderscore25 & 222.324 & 8.9485 & 0.55 &  3.5  $\pm$ 0.52 &  5.71 $\pm$ 1.07 & 11.58 $\pm$  2.03 &  8.01 $\pm$  2.46 &  7.19 $\pm$ 1.78 &  2.93 $\pm$ 2.07 & 1.75 $\pm$ 0.51 & Both \\
850\textunderscore31 & 222.292 & 8.9615 & 0.84 &  3.25 $\pm$ 0.52 &  8.44 $\pm$ 1.12 & 22.21 $\pm$  2.28 & 18.13 $\pm$  2.72 &  9.07 $\pm$ 1.92 & 11.05 $\pm$ 3.46 & 1.38 $\pm$ 0.5 & Both  \\
850\textunderscore16 & 222.289 & 9.0128 & 2.33 & 10.75 $\pm$ 0.95 & 22.18 $\pm$ 2.92 & 10.83 $\pm$  11.3 & 17.04 $\pm$ 12.56 &  3.24 $\pm$ 2.88 &  8.7  $\pm$ 6.94 & 1.58 $\pm$ 1.45 & Ez \\
850\textunderscore26 & 222.334 & 8.9446 & 0.8  &  2.28 $\pm$ 0.57 &  4.73 $\pm$ 1.31 &  1.6  $\pm$  1.61 &  3.3  $\pm$  3.36 &  1.54 $\pm$ 1.35 &  2.46 $\pm$ 2.47 & 0.93 $\pm$ 0.64 & Ez \\
850\textunderscore4 & 222.237 & 8.9284 & 2.23 &  0.39 $\pm$ 0.35 &  1.73 $\pm$ 1.03 &  6.55 $\pm$  2.52 & 10.6  $\pm$  3.89 &  7.88 $\pm$ 2.88 &  9.45 $\pm$ 4.46 & 5.48 $\pm$ 0.89 & Cg \\
850\textunderscore5 & 222.353 & 8.93   & 1.39 &  0.22 $\pm$ 0.23 &  3.17 $\pm$ 0.95 &  9.62 $\pm$  2.12 & 14.42 $\pm$  2.5  & 11.36 $\pm$ 2.35 & 15.63 $\pm$ 2.5  & 6.26 $\pm$ 0.8 & Cg  \\
850\textunderscore8 & 222.301 & 9.0196 & 2.48 &  2.05 $\pm$ 0.54 &  2.8  $\pm$ 1.09 &  9.2  $\pm$  2.13 & 12.06 $\pm$  2.87 &  9.18 $\pm$ 2.98 & 10.01 $\pm$ 4.26 & 7.03 $\pm$ 1.05 & Cg \\
850\textunderscore9 & 222.326 & 8.8754 & 2.03 &  1.86 $\pm$ 0.53 & 14.04 $\pm$ 1.25 & 26.55 $\pm$  2.34 & 34.25 $\pm$  3.04 &  6.75 $\pm$ 2.85 & 23.96 $\pm$ 2.69 & 6.4  $\pm$ 0.99 & Cg \\
850\textunderscore24 & 222.373 & 8.8732 & 2.83 &  1.58 $\pm$ 0.56 &  7.94 $\pm$ 1.23 & 24.82 $\pm$  2.36 & 31.12 $\pm$  2.9  &  8.17 $\pm$ 3.5  & 25.21 $\pm$ 2.75 & 3.73 $\pm$ 1 & Cg    \\
850\textunderscore1\textunderscore J & 222.306 & 8.9378 & 0.07 &  3.48 $\pm$ 0.56 & 10.98 $\pm$ 1.27 & 17.98 $\pm$  4.51 & 19.58 $\pm$  5.81 &  6.58 $\pm$ 1.91 & 12.57 $\pm$ 6.06 & 2.18 $\pm$ 0.64 & Cg \\
\hline
\end{tabular}}
\end{table*}

\subsection{Bright Cluster Core Sources}\label{sec:BCS}

\begin{figure}
\centering
\includegraphics[scale=0.45]{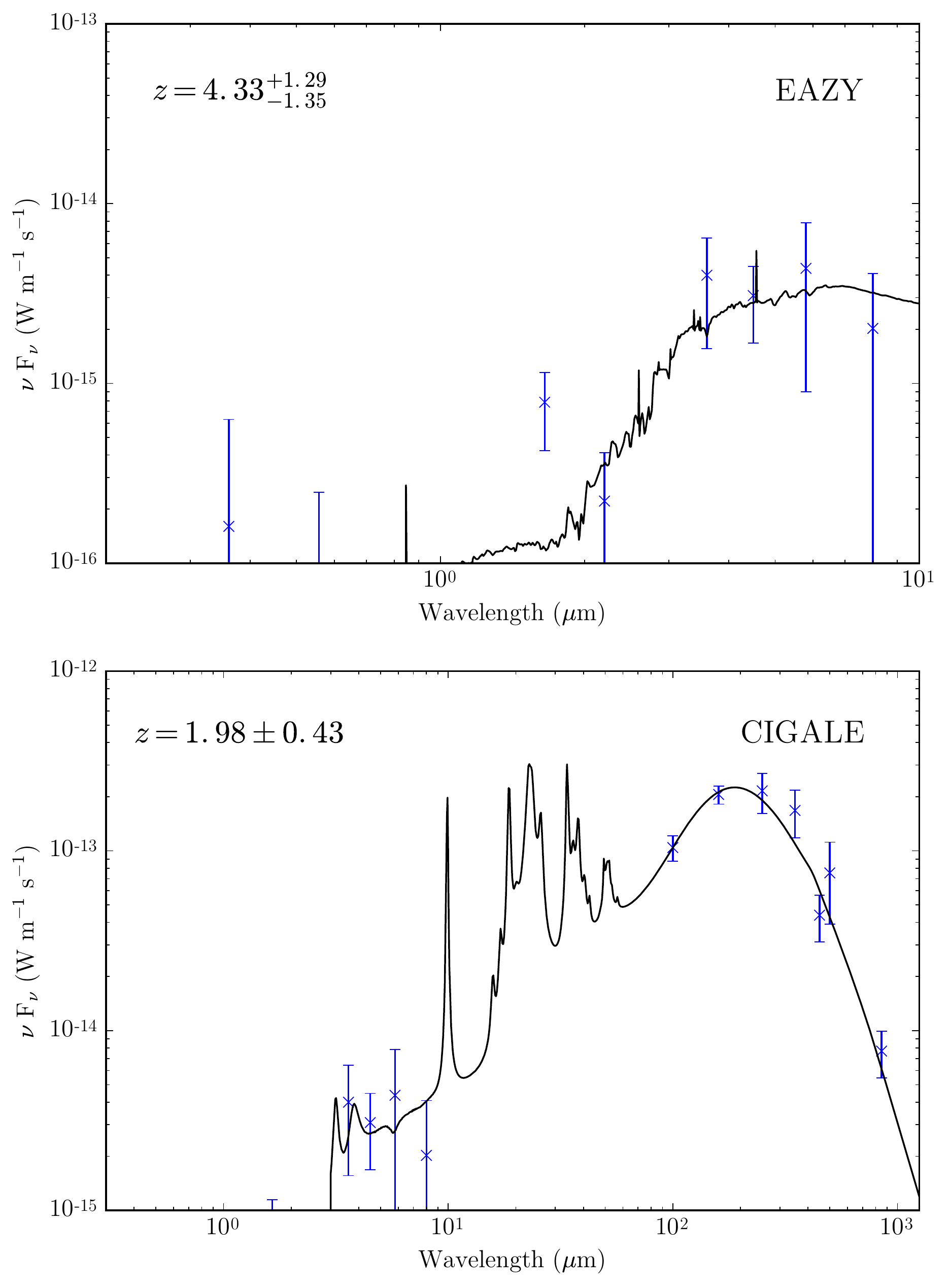}
\caption{Comparing the SEDs from both \textsc{EAZY} and \textsc{CIGALE} for source 850\textunderscore1\textunderscore J. Notice that by including the sub-mm data the source is placed within the cluster}\label{fig:A4}
\end{figure}

In \citetalias{2018Coogan} no spectral redshifts were detected for sources 850\textunderscore1\textunderscore I and 850\textunderscore1\textunderscore J (A5 and A4 in \citetalias{2018Coogan}). When we calculated photometric redshifts for them we found that 850\textunderscore1\textunderscore I was not placed in the cluster for either method, both indicating it had a redshift of $\sim$2.8. When we look at the other source, 850\textunderscore1\textunderscore J we see that indeed \textsc{EAZY} placed it well outside of the cluster (z$\sim$4.3) whereas \textsc{CIGALE} placed it within the cluster (Figure \ref{fig:A4}).  

Combined 850\textunderscore1\textunderscore I and 850\textunderscore1\textunderscore J contribute 70\% of the 870\,$\mu$m flux for the cluster core, meaning that the chance of these galaxies being interlopers is extremely small. Based on 870\,$\mu$m source counts from \cite{2013Karim} the chance of not being associated with the cluster is less than 4$\times$10$^{-5}$ (\citetalias{2018Coogan}). Lensing is an unlikely cause to the high fluxes simply because the mass of the cluster halo is not enough to boost the fluxes to the observed levels. A full discussion on this is presented in both \citetalias{2018Coogan} and \citetalias{2018Strazzullo}, who exclude both sources from there analysis. For the rest of our analysis we include 850\textunderscore1\textunderscore J as a potential member based on its \textsc{CIGALE} redshift. When appropriate we also consider the situation that this galaxy is not in the cluster. We should note that the errors on 850\textunderscore1\textunderscore I are substantial for both \textsc{EAZY} and \textsc{CIGALE}, with both having $\Delta z > 1$. Whilst there is a chance it could be in the cluster the large error means we cannot say for sure, and therefore the source is excluded from the rest of the analysis.

\section{Star Forming Properties Of CLJ1449}\label{sec:SFR}

With our list of cluster members we used \textsc{CIGALE} to fit SEDs for our analysis. Again the same settings as above were used, this time however the redshift was fixed at 2 for all galaxies. We used \textsc{CIGALE} rather then conventional methods of fitting a Modified Black-Body (MBB), because as mentioned the uncertainties on the FIR/sub-mm data are large, and we obtained better constraints using \textsc{CIGALE}. We find SFRs between 20-1600\,M$_{\odot}$ yr$^{-1}$ with a median value of 168\,M$_{\odot}$ yr$^{-1}$. The results from \textsc{CIGALE} can be found in Table \ref{table:properties}, and all optical and NIR data can be found in Table \ref{table:optproperties} and \ref{table:MIRproperties}.

\begin{table*}
\caption{FIR/sub-mm properties of our cluster galaxies.}  
\label{table:properties}     
\small
\centering           
\begin{tabular}{ccccccccccccc} 
\hline\hline   
ID  & $z_{Spec}$ &  $z_{EZ}$ & $z_{CG}$&  $L_{IR}$ &  $SFR$ & $M_{*}$  & $P$ \\
&&&&(10$^{12}$\,L$_{\odot}$)&(M$_{\odot}$yr$^{-1}$)&(10$^{11}$\,M$_{\odot}$) & \\\hline                         
850\textunderscore1\textunderscore A & 1.9951 $\pm$ 0.0004 & 2.33$^{+0.34}_{-0.31}$ & 2.25 $\pm$ 0.25 &  0.21 $\pm$ 0.15 &   32 $\pm$  17 & 0.16 $\pm$ 0.08  & - \\
850\textunderscore1\textunderscore B & 1.9902 $\pm$ 0.0005 & 2.25$^{+0.53}_{-0.55}$ & 2.14 $\pm$ 0.37 &  0.18 $\pm$ 0.22 &   22 $\pm$  28 & 0.17 $\pm$ 0.21  & - \\
850\textunderscore1\textunderscore C & 1.9944 $\pm$ 0.0006 & 2.51$^{+0.64}_{-0.87}$ & 1.92 $\pm$ 0.52 &  0.62 $\pm$ 0.38 &   74 $\pm$  59 & 0.36 $\pm$ 0.17  & - \\
850\textunderscore1\textunderscore D & 1.9832 $\pm$ 0.0007 & 2.35$^{+0.32}_{-0.28}$ & 1.70 $\pm$ 0.54 &  0.38 $\pm$ 0.10 &   36 $\pm$  13 & 0.43 $\pm$ 0.13  & - \\
850\textunderscore1\textunderscore E & 1.9965 $\pm$ 0.0004 & 3.45$^{+0.79}_{-0.70}$ & 2.23 $\pm$ 1.03 &  0.52 $\pm$ 0.48 &   66 $\pm$  69 & 0.26 $\pm$ 0.37  & - \\
850\textunderscore1\textunderscore F & 1.9883 $\pm$ 0.0070 & 1.64$^{+0.62}_{-0.66}$ & 2.02 $\pm$ 0.30 &  0.37 $\pm$ 0.30 &   45 $\pm$  49 & 0.41 $\pm$ 0.22  & - \\
850\textunderscore1\textunderscore G & 1.9903  $\pm$0.0004 & 1.82$^{+0.39}_{-0.47}$ & 1.59 $\pm$ 0.37 &  0.41 $\pm$ 0.37 &   59 $\pm$  67 & 0.30 $\pm$ 0.29  & - \\
850\textunderscore1\textunderscore H & 1.982 $\pm$ 0.002 & 2.43$^{+0.24}_{-0.24}$ & 3.62 $\pm$ 0.60 &  0.56 $\pm$ 0.67 &   73 $\pm$ 100 & 0.26 $\pm$ 0.44  & - \\
850\textunderscore1\textunderscore K & 1.98 $\pm$ 0.02 & 1.24$^{+1.20}_{-0.83}$ & 1.63 $\pm$ 0.58 &  0.30 $\pm$ 0.11 &   37 $\pm$  16 & 0.30 $\pm$ 0.12  & - \\
850\textunderscore13 & - & 1.98$^{+0.44}_{-0.47}$ & 2.25 $\pm$ 0.24 &  0.94 $\pm$ 0.23 &  116 $\pm$  22 & 2.08 $\pm$ 0.53 & 0.04 \\
850\textunderscore14 & - & 2.10$^{+0.42}_{-0.47}$ & 1.79 $\pm$ 0.24 &  3.03 $\pm$ 0.36 &  331 $\pm$  88 & 4.09 $\pm$ 0.99  & 0.01\\
850\textunderscore17 & - & 2.34$^{+0.87}_{-0.93}$ & 1.98 $\pm$ 0.40 &  3.65 $\pm$ 1.16 &  447 $\pm$ 156 & 3.86 $\pm$ 1.18  & 0.04\\
850\textunderscore20 & - & 2.07$^{+0.52}_{-0.51}$ & 2.04 $\pm$ 0.21 &  4.98 $\pm$ 0.38 &  439 $\pm$  98 & 6.84 $\pm$ 1.48  & 0.08\\
850\textunderscore22 & - & 2.35$^{+0.43}_{-0.46}$ & 2.04 $\pm$ 0.32 &  3.30 $\pm$ 1.85 &  403 $\pm$ 214 & 2.28 $\pm$ 1.12  & 0.01\\
850\textunderscore25 & - & 1.89$^{+0.22}_{-0.26}$ & 2.06 $\pm$ 0.27 &  2.46 $\pm$ 0.44 &  254 $\pm$  71 & 2.99 $\pm$ 0.83  & 0.01\\
850\textunderscore31 & - & 2.18$^{+0.77}_{-0.68}$ & 2.17 $\pm$ 0.32 &  2.73 $\pm$ 0.41 &  377 $\pm$ 117 & 1.33 $\pm$ 0.56  & 0.08\\
850\textunderscore16 & - & 2.34$^{+0.58}_{-0.57}$ & 1.56 $\pm$ 0.21 & 11.64 $\pm$ 2.71 & 1618 $\pm$ 738 & 7.25 $\pm$ 6.19 & 0.01\\
850\textunderscore26 & - & 1.90$^{+1.14}_{-0.86}$ & 1.26 $\pm$ 0.36 &  2.35 $\pm$ 0.98 &  313 $\pm$ 160 & 0.74 $\pm$ 0.52 & 0.09\\
850\textunderscore4 & - & 1.56$^{+0.92}_{-0.97}$ & 1.96 $\pm$ 0.34 &  0.93 $\pm$ 0.18 &  105 $\pm$  30 & 0.58 $\pm$ 0.39 & 0.002\\
850\textunderscore5 & - & 1.46$^{+0.87}_{-0.78}$ & 1.98 $\pm$ 0.70 &  1.26 $\pm$ 0.14 &  142 $\pm$  40 & 0.61 $\pm$ 0.32 & 0.004\\
850\textunderscore8 & - & 1.58$^{+0.49}_{-0.41}$ & 2.10 $\pm$ 0.97 &  1.71 $\pm$ 0.28 &  195 $\pm$  63 & 0.74 $\pm$ 0.49  & 0.01\\
850\textunderscore9 & - & 3.52$^{+0.35}_{-0.36}$ & 2.07 $\pm$ 0.39 &  3.55 $\pm$ 0.27 &  357 $\pm$  86 & 2.51 $\pm$ 1.22  & 0.002\\
850\textunderscore24 & - & 2.61$^{+0.60}_{-0.56}$ & 2.27 $\pm$ 0.39 &  2.99 $\pm$ 0.34 &  360 $\pm$ 101 & 1.67 $\pm$ 0.64  & 0.01\\
850\textunderscore1\textunderscore J & - & 4.33$^{+1.29}_{-1.35}$ & 1.98 $\pm$ 0.43 &  2.98 $\pm$ 0.28 &  381 $\pm$ 138 & 1.31 $\pm$ 1.33  & - \\
\hline
\end{tabular}
\end{table*}

\subsection{Radial Variations In The Star Formation Rate Density}

\begin{figure*}
\centering
\includegraphics[scale=0.8]{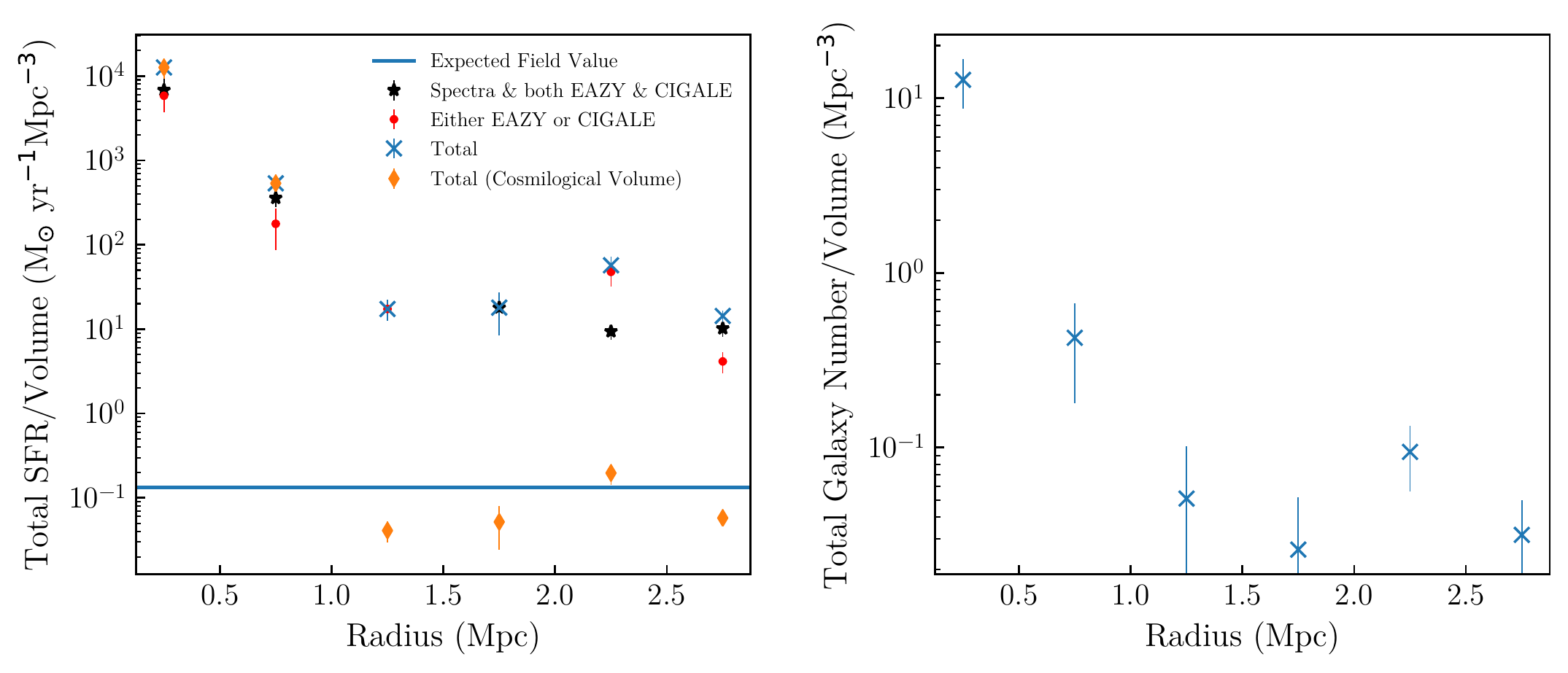}
\caption{Left: The SF-density relation found in CLJ1449. We have divided the cluster into 0.5\,Mpc bins, normalised by the volume of the bin. Each bin has been separated by the redshift criteria mentioned in Section \ref{sec:RS}. The blue line is the expected field value based on \citet{2014Madau}. We also plot the SF-density relation by assuming the observed cosmological volume between 1.6\,$<$z\,$<$2.4. Right: The number density of galaxies versus the cluster radius. For sources less than 1\,Mpc the counts are from ALMA, whilst at larger radii we use radio counts}\label{fig:SFRV}
\end{figure*}

To understand the radial variation in the SFR of the cluster, we calculated the distance from the centre for each galaxy and binned them in 0.5\,Mpc bins ranging from 0-3\,Mpc. We summed up the SFRs in each radial bin, and normalise it by the volume of the bin. The resulting plot can be seen in Figure \ref{fig:SFRV}.

We confirm that the central 0.5\,Mpc region is highly star forming with a total SFR of 800\,$\pm$\,200\,M$_{\odot}$ yr$^{-1}$. When converted into a SFRD we find a projected volume density of (1.2\,$\pm$\,0.3)$\times$10$^{4}$\,M$_{\odot}$yr$^{-1}$Mpc$^{-3}$, which is almost five orders of magnitude greater than the expected value for field galaxies ($\sim$0.1\,M$_{\odot}$ yr$^{-1}$Mpc$^{-3}$, \citealt{2014Madau}). With increasing radius we see a decrease in the SFRD of almost 2 orders of magnitude, until it stabilizes at 1\,Mpc, where it remains constant with the exception of a spike between 2-2.5\,Mpc. A similar spike was seen in \cite{2014Santos} in a redshift 1.6 cluster, and was caused by an increase in the number of high mass galaxies at this radius. This is again seen in CLJ1449 with the heaviest galaxy in our sample (850\textunderscore16, M$_{*}\sim$7.5 $\times$10$^{11}$\,M$_{\odot}$) being found in this radius bin.

For completeness we also considered the scenario where 850\textunderscore1\textunderscore J is \textit{not} associated with the cluster (the black star in the 0-0.5\,Mpc bin in Figure \ref{fig:SFRV}). We find that the SFR for the central 0.5\,Mpc region decreases to 440\,$\pm$\,160\,M$_{\odot}$ yr$^{-1}$ a factor of $\sim$\,2 lower then before. Again when converted to a SFRD this gives (6.7\,$\pm$\,2.4)$\times$10$^{3}$\,M$_{\odot}$yr$^{-1}$Mpc$^{-3}$, which whilst lower then before, it is still much larger than the expected result from \cite{2014Madau} for field galaxies.

In Figure \ref{fig:SFRV} we also plot the number density of sources as a function of radius. Again we see an identical trend to what is observed with the SFR-density relation. This would indicate that the observed reversal in the SF-density relation is caused by a dense population of SF galaxies, rather then a small population of extreme starbursts.

We stress that the high number counts in the central 1\,Mpc region is down to the high resolution ALMA data, whereas those counts at radii greater than 1\,Mpc are based on radio sources. If we base our number counts entirely on the radio data then we find that the number density in the central region reduces by half. Whilst this is still a significant, it could indicate that we are not observing all the galaxies within the cluster.

The biggest issue with the galaxies beyond 1\,Mpc is the redshift and their uncertainties. When doing the \textsc{CIGALE} fitting we fix the redshift to that of the cluster, so if we have an interloper this could cause an increase in the SFR. To test for this we normalise our SFR with the cosmological volume between 1.6\,$<$z\,$<$2.4 for radii greater than 1\,Mpc (as opposed to the cluster volume). When accounting for just the cosmological volume (Figure \ref{fig:SFRV}) the SF-density does indeed fall to expected field values at radii greater than 1\,Mpc. This indicates that there is a high chance of significant contamination in our sample, which causes an artificial increase in the SF.

We do note that the two volumes we have normalised by are the two extreme cases. Using the cluster volume assumes all sources above 1\,Mpc are in the cluster, and the cosmological volume assumes that none are. Whilst we do acknowledge that either one of these extremes could be occurring (no matter how unlikely), the fact that even at a radius greater than 1\,Mpc we are still in a very over-dense region ($\sim$\,2\,$\times$ over-dense) the chances are we are in the middle of these scenarios. This means we still expect elevation in the SF-density relation above the expected value, but it will not be as extreme as seen in Figure \ref{fig:SFRV}.

\subsection{Star Formation Rate vs Stellar Mass}

\begin{figure*}
\centering
\includegraphics[scale=0.8]{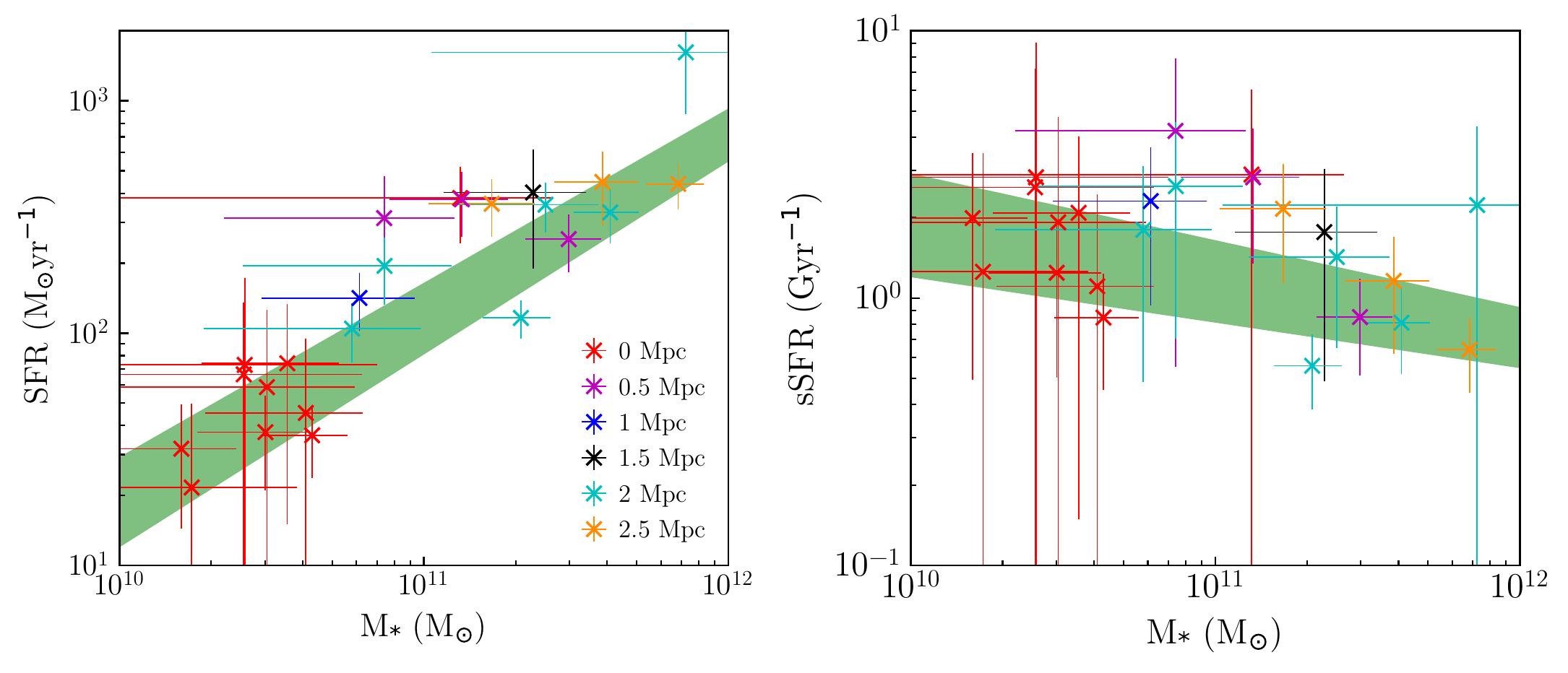}
\caption{Left: The SFR-M$_{*}$ relation with the shaded region showing the expected position for main sequence galaxies at redshift 2 given by \citet{2014Sargent}. We have divided the points up into there location within the cluster and colour coded them accordingly. Right: The sSFR-M$_{*}$ relation with the expected value for redshift 2 main sequence galaxies.}\label{fig:sSFR}
\end{figure*}

We then investigated the relation between the stellar masses (M$_{*}$) of our cluster galaxies and their SFRs. Using stellar masses from \textsc{CIGALE} we compare them to the SFRs, with the resulting relation shown in Figure \ref{fig:sSFR}. We compare these galaxies to the expected galaxy Main Sequence (MS) relation for redshift 2 galaxies from \cite{2014Sargent}. Even though all these galaxies are highly star forming, and are extremely luminous (with all of them being LIRGs or ULIRGs) we find that the majority of galaxies within the cluster do not deviate from the expected main sequence. This behaviour was observed by \citetalias{2018Coogan} when observing the central galaxies. However they noted that these galaxies had star-burst like behaviour, based on the gas excitation.\\
\indent Figure \ref{fig:sSFR} would seem to suggest that the environment does not have a significant effect on the SF properties of these galaxies. \cite{2016Darvish} found that in the COSMOS field, galaxies at low redshift (z\,$<$\,1) seemed to be heavily influenced by the environment they reside in, and in dense environments the SFR decreased significantly. However at z\,$>$\,1 the SFR did not significantly change with environment, suggesting that at high redshift the SFR-M$_{*}$ relation is independent of environment. This behaviour was also seen by \cite{2013Koyama} who again suggest the SFR-M$_{*}$ relation is not dependant on the galaxy environment at high redshift.\\
\indent We can link the low level of starbursts back to the density of sources seen in Figure \ref{fig:SFRV}. If we had a low number density of sources in the cluster core, we would expect to see far more galaxies exhibiting star burst behaviour to account for the large SF-density. The fact that we have a high number density of sources in the cluster core and limited starbursts reinforce the claim that the SFR-density relation is caused by a high density of SF galaxies sources rather than a small population of star bursting galaxies.\\
\indent A small sample of our galaxies do appear to lie above the MS, being $\sim$2-3\,$\times$ above the expected relation. This enhanced SF activity could indeed be down to merging events, as both \cite{2013Hung} and \cite{2018Cibinel} found that galaxies undergoing a merging event lie above the galaxy MS. This would also validate claims made by \citetalias{2018Coogan} that most of the activity in this cluster is driven by mergers. It should be noted that these galaxies that are above the main sequence are those galaxies that had their redshifts estimated by either \textsc{EAZY} \textit{or} \textsc{CIGALE}. As mentioned earlier this increase in SF could be an artifact of fixing the redshift when running \textsc{CIGALE}.\\
\indent Mergers could also explain why there is elevated SF (compared to the field) well beyond the viral radius of the cluster. Within the cluster core itself, galaxies are moving too fast for a merger to actually occur.  However in the in-fall region the speed of galaxies is much lower allowing mergers to occur, and then be accreted onto the core (\citealt{1998Bekki}, \citealt{2006Moss}). Whilst measuring velocity dispersion is still very difficult at high redshift, \citealt{2017Delahaye} showed evidence that the fraction of mergers in a redshift 1.6 cluster core is no higher than field values, indicating that the mergers are still happening in the outskirt rather than in the core itself. So in the outskirts of the clusters, mergers can occur more frequently and cause an increase in star formation. Whilst signs of merger activity cannot be detected with the current ground based optical data, it is hoped with new high resolution data this can be investigated further.\\
\indent Whilst we have discussed the fact that environment and SF are independent at high redshift, it should be noted that in the left panel of Figure \ref{fig:sSFR} the galaxies with the smallest cluster radii have the lowest stellar masses. This could suggest that the cluster environment is beginning to have an impact on the cluster galaxies, and quenching is starting to occur within these galaxies. Similar conclusions were drawn by both \citetalias{2018Coogan} observing this cluster, and \citealt{2014Santos} observing a redshift 1.6 cluster.\\
\indent In Figure \ref{fig:sSFR} we convert our SFR to a specific SFR (sSFR, SFR/M$_{*}$) and observe a large scatter from the proposed relation outlined in \cite{2014Sargent}. Because the sSFR measures how the current SFR compares to to the SFR of the galaxy averaged over its life, the fact these galaxies have heightened sSFRs indicates that they are undergoing bursts of star formation.\\
\indent A decrease in sSFR with increasing stellar mass has been reported at all redshifts (e.g. \citealt{2015Ilbert}, \citealt{2015Lehnert}) and is believed to show that the most massive galaxies have already formed all their stellar mass much earlier then their lighter counterparts. Whilst with this data suggests that there is a trend, it is not statistically significant enough to be considered, with a Spearman's rank correlation coefficient of only -0.3. This could indicate that the heaviest galaxies are still in the process of forming within the cluster.

\subsection{Mass Normalised Star Formation Rate}\label{sec:MNSFR}

As we have discussed there is strong indication for an increase in the SF activity in clusters with increasing redshift. It has been shown that CLJ1449 is still very actively star forming but is this in line with what would be expected at this redshift. To allow for direct comparisons to other studies we follow the same methodology outlined in \cite{2012Popesso}. We first integrate out to a radius of 1\,Mpc (as in studies such as \citealt{2014Santos} and \citealt{2015Ma}), giving a total integrated SFR ($\Sigma$SFR) of 1800\,$\pm$\,300\,M$_{\odot}$yr$^{-1}$. We then normalise $\Sigma$SFR by the cluster mass ((0.5$\pm$0.1) $\times$10$^{14}$\,M$_{\odot}$) resulting in a normalised, $\Sigma$SFR/M$_{cl}$ of 3300\,$\pm$\,850\,M$_{\odot}$ yr$^{-1}$/10$^{14}$ M$_{\odot}$.

\begin{figure}
\centering
\includegraphics[scale=0.7]{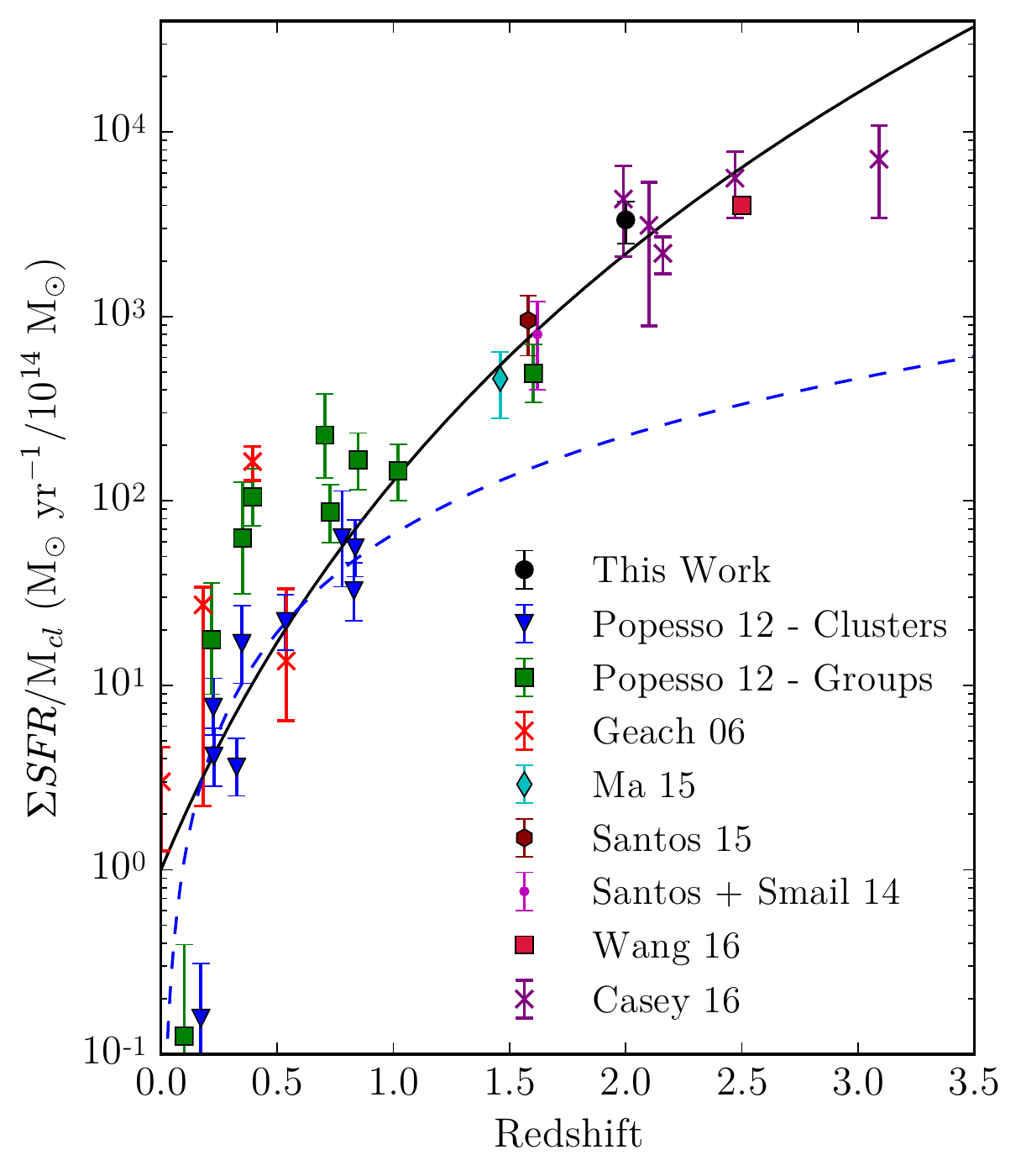}
\caption{Comparison of the mass normalised $\Sigma$SFR for several clusters at different redshifts. The solid black line is the proposed relation offered by \citet{2004Cowie} and \citet{2006Geach} that follows the relation $(1+z)^{7}$, and the blue dashed line follows the relation in \citet{2012Popesso} for clusters. It should be noted that the sample from \citet{2016Casey} are proto-clusters and not fully virialized clusters.}\label{fig:Mass_sfr}
\end{figure}

In Figure \ref{fig:Mass_sfr} we compare our result to clusters at several different redshifts. We see that our cluster lies above the expected evolutionary trend predicted by Popesso, however this cluster is at a much higher redshift then their sample. When comparing the theorised (1+$z$)$^{7}$ line  in \cite{2004Cowie} and \cite{2006Geach}, to CLJ1449 we see that there is only a 1.3\,$\sigma$ offset, meaning that CLJ1449 could follow trends seen at both low and high redshift. When comparing to other high redshift (proto-)clusters from both \cite{2016Wang} (CLJ1001) and \cite{2016Casey} we see that there is significant scatter. Other studies of proto-clusters such as those by \cite{2014Shimakawa} and \cite{2018Lacaille}  has also showed significant scatter at z$\,>\,$2. This scatter was observed by \cite{2006Geach} at low redshift and suggests that the cluster environments have differing, but strong influences on the star formation histories of the residing galaxies. This is very significant for the sample from \cite{2016Casey} which are proto-clusters, and are still in the process of forming.

It should also be noted that both CLJ1449 and CLJ1001 are significantly less massive then other clusters used in previous studies, both at low and high redshift. The fact that these clusters do not directly fit to the $(1+z)^7$ relation, and there is some scatter that could indicate cluster mass could have some influence on the galaxies within them. This was also suggested by both \cite{2011Koyama} and \cite{2018Cochrane} who observed similar scatter in z$\,>\,$1 clusters. Cluster mass and the SF properties of cluster galaxies will be investigated in future works. However for the time being high redshift samples are still too small to determine if all clusters have enhanced SF activity, or do they behave like low redshift galaxies and we have a biased sample.

\section{Conclusions}

We have presented new SCUBA-2 images of CLJ1449, a mature cluster at redshift two. We combine this data with pre-existing data including \textit{Herschel}, ALMA and VLA to study the SF properties of this cluster, and build upon the work already presented in \citetalias{2018Coogan} and \citetalias{2018Strazzullo}.

\begin{itemize}
\item We use SCUBA-2 and \textit{Herschel} data to explore $\sim$0.03\,deg$^2$ that contain the cluster CLJ1449. We identify 32 sources in the 850\,$\mu$m maps that have a S/N greater then 4. 
\item To help estimate fluxes for confused members we use high resolution ALMA and JVLA maps to identify positions of potential FIR/sub-mm galaxies. We then use the Bayesian inference tool XID+ to estimate fluxes for all these sources in the confused \textit{Herschel} and SCUBA-2 maps.
\item We match these sources up to our SCUBA-2 sources resulting in 37 potential cluster members. To confirm cluster membership we calculate redshifts using both \textsc{EAZY} and \textsc{CIGALE}. We find 24 galaxies we are confident could be within the cluster.
\item We use \textsc{CIGALE} estimates for both SFRs and stellar masses for all 24 galaxies. We find that the central 0.5\,Mpc region is very highly star forming, forming 800$\pm$200\,M$_{\odot}$yr$^{-1}$, which corresponds to a SFRD of (1.2\,$\pm$\,0.3)$\times$10$^{4}$\,M$_{\odot}$yr$^{-1}$Mpc$^{-3}$. This is orders of magnitude greater then field values.
\item When looking at the SFR-M$_{*}$ relation we see these galaxies lie on the expected main sequence relation, however there is some evidence of star-bursting activity which could possibly be caused by merger events. When looking at the sSFR-M$_{*}$ we see a large scatter in this relation which could indicate that the gas is undergoing complex processes.
\item When comparing the mass normalised integrated SFR we see that CLJ1449 seems to follow previously identified scaling relations (with minimal scatter), but there is still a large scatter when considering other high redshift systems. However due to the low number of high redshift systems it is unknown if this is reflective of what is seen in low redshift systems.   
\end{itemize}

\section*{Acknowledgements}
We thank the referee for their extremely useful comments and suggestions. We also thank Rapha\"{e}l Gobat for providing the optical and NIR maps of the cluster and Thomas Williams for providing the SDSS image.

CMAS acknowledges support from the UK Science  and  Technology  Facilities  Council.

MWLS acknowledge  funding  from  the  UK  Science  and  Technology  Facilities  Council  consolidated  grant  ST/K000926/1. MWLS and SAE have also received funding from the European Union Seventh Framework Programme ([FP7/2007-2013] [FP7/2007-2011]) under grant agreement no. 607254.

The  James  Clerk  Maxwell  Telescope  is operated by the East Asian Observatory on behalf of The National Astronomical Observatory of Japan, Academia Sinica Institute of Astronomy and Astrophysics, the Korea Astronomy and Space Science Institute, the National Astronomical Observatories of China and the Chinese Academy of Sciences (grant no. XDB09000000), with additional funding support from the Science and Technology
Facilities Council of the United Kingdom and participating universities in the United Kingdom and Canada.

PACS  has  been  developed  by  a  consortium  of  institutes  led by  MPE  (Germany)  and  including  UVIE  (Austria);  KUL,  CSL, IMEC (Belgium); CEA, OAMP (France); MPIA (Germany); IFSI, OAP/AOT, OAA/CAISMI, LENS, SISSA (Italy); and IAC (Spain). This development has been supported by the funding   agencies  BMVIT  (Austria),  ESA-PRODEX  (Belgium),  CEA/CNES (France), DLR (Germany), ASI (Italy) and CICYT/MCYT (Spain).

SPIRE has been developed by a consortium of institutes led by Cardiff Univ. (UK) and including Univ. Lethbridge (Canada); NAOC  (China);  CEA,  LAM  (France);  IFSI,  Univ.  Padua  (Italy); IAC (Spain); Stockholm Observatory (Sweden); Imperial College London,  RAL,  UCL-MSSL,  UK  ATC,  Univ.  Sussex  (UK) and Caltech,  JPL,  NHSC,  Univ.  Colorado  (USA).  This  development has been supported by national funding agencies: CSA (Canada); NAOC (China); CEA, CNES, CNRS (France); ASI (Italy); MCINN (Spain);  SNSB (Sweden);  STFC  and  UKSA  (UK);  and  NASA (USA).

This paper makes use of ALMA data 2012.1.00885.S. ALMA is a partnership of ESO (representing its member states), NSF (USA) and NINS (Japan),together with NRC (Canada), MOST and ASIAA  (Taiwan),and KASI  (Republic of Korea), in cooperation with the Republic of Chile. The Joint ALMA Observatory is operated by ESO,AUI/NRAO and NAOJ.

This paper also makes use of JVLA program 12A-188. The Na-tional Radio Astronomy Observatory is a facility of the NationalScience Foundation operated under cooperative agreement by As-sociated Universities, Inc.

This research made use of Astropy, a community-developed core Python package for Astronomy (\citealt{2013Astropy}), and  Matplotlib,  a  Python  2D  plotting  library  (\citealt{2007Hunter}).





\bibliographystyle{mnras}
\bibliography{The_Paper}



\appendix

\section{Optical and NIR Data For Cluster Members}

\begin{table*}
\caption{Optical and NIR data of our cluster galaxies. The units are in $\mu$Jy.}  
\label{table:optproperties}     
\small
\centering           
\begin{tabular}{ccccccc} 
\hline\hline   
ID  & U & B & V & I & Z & Y  \\ \hline                         
850\textunderscore1\textunderscore A & 0.45 $\pm$ 0.28 & 1.32 $\pm$ 0.2 & 1.25 $\pm$ 0.31 & 0.07 $\pm$ 0.22 & 1.15 $\pm$ 0.23 & 0.75 $\pm$	0.41 \\
850\textunderscore1\textunderscore B & 0.06 $\pm$ 0.1 &	0.14 $\pm$ 0.02 & 0.1 $\pm$	0.07 & 0.21 $\pm$	0.04 & 0.2 $\pm$ 0.04 &	0.16 $\pm$ 0.18 \\
850\textunderscore1\textunderscore C & 0.11 $\pm$ 0.14 & 0.06 $\pm$	0.01 & 0.13 $\pm$ 0.08 & 0.33 $\pm$	0.07 & 0.4 $\pm$ 0.08 &	0.33 $\pm$ 0.26 \\
850\textunderscore1\textunderscore D & 0.01	$\pm$ 0.05 & - & - & - &	0.06 $\pm$	0.01 &	0.22 $\pm$	0.21 \\
850\textunderscore1\textunderscore E & 0.02 $\pm$	0.06 &	- &	0.08 $\pm$	0.06 &	0.13 $\pm$	0.03 &	0.19 $\pm$	0.04 &	0.07 $\pm$	0.12 	  \\
850\textunderscore1\textunderscore F & 0.18 $\pm$	0.18 &	0.33 $\pm$	0.05 &	0.21 $\pm$	0.1	& 0.54 $\pm$	0.11 &	0.77 $\pm$	0.15 &	0.56 $\pm$	0.34  \\
850\textunderscore1\textunderscore G & 0.06 $\pm$	0.1 &	0.16 $\pm$	0.02 &	- &	0.14 $\pm$	0.03 &	0.54 $\pm$	0.11 &	-  \\
850\textunderscore1\textunderscore H & 0.02 $\pm$	0.05 &	0.09 $\pm$	0.01 &	0.05 $\pm$	0.05 &	- &	0.05 $\pm$	0.01 &	0.1 $\pm$	0.14  \\
850\textunderscore1\textunderscore K & - &	0.25 $\pm$	0.04 &	- &	0.72 $\pm$	0.14 &	2.16 $\pm$	0.43 &	0.7 $\pm$	0.39 \\
850\textunderscore13 & -	 & 3.19	 $\pm$0.48 &	- &	2.81 $\pm$	0.56 &	4.57 $\pm$	0.91 &	- \\
850\textunderscore14 & 1.73 $\pm$	0.59 &	2.61 $\pm$	0.4 &	2.16 $\pm$	0.45 &	2.87 $\pm$	0.57 &	4.99 $\pm$	1.0 &	- \\
850\textunderscore17 & - &	2.98 $\pm$	0.45 & - &	8.14 $\pm$	1.63 &	12.96 $\pm$	2.59 &	- \\
850\textunderscore20 & - &	0.35 $\pm$	0.05 &	- &	0.8 $\pm$	0.16 &	1.45 $\pm$	0.29 &	-  \\
850\textunderscore22 & - &	7.32 $\pm$	1.1 &	- &	5.84 $\pm$	1.17 &	8.24 $\pm$	1.65 &	- \\
850\textunderscore25 & 0.41 $\pm$	0.27 &	0.89 $\pm$	0.13 &	0.97 $\pm$	0.26 &	1.96 $\pm$	0.39 &	3.02 $\pm$	0.6 &	4.33 $\pm$	1.25  \\
850\textunderscore31 & - &	0.08 $\pm$	0.01 &	0.13 $\pm$	0.08 &	- &	0.68 $\pm$	0.14 &	0.46 $\pm$	0.31 \\
850\textunderscore16 & - &	2.08 $\pm$	0.31 &	- &	2.89 $\pm$	0.58 &	5.13 $\pm$	1.03 &	-  \\
850\textunderscore26 & - &	0.56 $\pm$	0.08 &	- &	1.18 $\pm$	0.24 &	2.56 $\pm$	0.51 &	- \\
850\textunderscore4 & - &	0.81 $\pm$	0.12 &	- &	0.87 $\pm$	0.17 &	1.05 $\pm$	0.21 &	- \\
850\textunderscore5 & - &	0.42 $\pm$	0.06 &	-	 &0.49 $\pm$	0.1 &	0.76 $\pm$	0.15 &	-  \\
850\textunderscore8 & - &	0.46 $\pm$	0.07 & 	- &	0.33 $\pm$	0.07 &	0.95 $\pm$	0.19 &	-  \\
850\textunderscore9 & 0.04 $\pm$	0.08 &	0.04 $\pm$	0.01 &	0.2 $\pm$	0.1 &	0.47 $\pm$	0.09 &	0.84 $\pm$	0.17 &	0.41 $\pm$	0.29  \\
850\textunderscore24 & 0.08 $\pm$	0.12 &	0.43 $\pm$	0.07 &	- & 1.16 $\pm$	0.23 &	1.43 $\pm$	0.29 &	-	\\
850\textunderscore1\textunderscore J & 0.02 $\pm$	0.06 &	-	 & 0.02 $\pm$	0.03 &	- &	- &	-  \\
\hline
\end{tabular}
\end{table*}

\begin{table*}
\caption{ NIR and MIR data for our cluster galaxies. The units are in $\mu$Jy.}  
\label{table:MIRproperties}     
\small
\centering           
\begin{tabular}{cccccccc} 
\hline\hline   
ID & J & H & K & 3.6\,$\mu$m & 4.8\,$\mu$m & 5.8\,$\mu$m & 8\,$\mu$m\\ \hline          

850\textunderscore1\textunderscore A &1.68 $\pm$	0.6 & 2.57 $\pm$ 0.68 &	3.72 $\pm$ 0.98 & 6.92 $\pm$ 4.34 & 9.59 $\pm$ 5.05 &	6.16 $\pm$ 6.36 & 6.77 $\pm$ 6.26 \\
850\textunderscore1\textunderscore B &0.16 $\pm$ 0.18 & 0.56 $\pm$ 0.31 &	1.08 $\pm$ 0.36 & 2.94 $\pm$ 0.82 &	26.35 $\pm$ 16.39 & - & -   \\
850\textunderscore1\textunderscore C &1.15 $\pm$ 0.47 &	2.72 $\pm$ 0.71 & 3.39 $\pm$ 0.92 &	4.66 $\pm$ 3.15 & 6.72 $\pm$	3.92 & 15.28 $\pm$ 12.91 & 15.59 $\pm$ 8.21 \\
850\textunderscore1\textunderscore D &1.19 $\pm$	0.48 &	2.54 $\pm$	0.67 &	4.41 $\pm$	1.13 &	9.31 $\pm$	5.33 &	6.25 $\pm$	4.27 &	14.27 $\pm$	8.04 &	1.43 $\pm$	4.74 \\
850\textunderscore1\textunderscore E &0.15 $\pm$	0.15 &	1.05 $\pm$	0.35 &	1.65 $\pm$	0.54 &	3.88 $\pm$	2.81 &	11.54 $\pm$	8.79 &	12.59 $\pm$	8.08 &	23.9 $\pm$	10.32  \\
850\textunderscore1\textunderscore F &2.19 $\pm$	0.72 &	3.67 $\pm$	0.9 &	7.27 $\pm$	1.71 &	- &	- &	- &	- \\
850\textunderscore1\textunderscore G &1.1 $\pm$	0.46 &	2.51 $\pm$	0.66 &	2.58 $\pm$	0.75 &	12.71 $\pm$	7.9 &	21.75 $\pm$	14.4 &	18.86 $\pm$	10.62 &	18.39 $\pm$	8.9  \\
850\textunderscore1\textunderscore H &0.31 $\pm$	0.22 &	1.13 $\pm$	0.37 &	1.84 $\pm$	0.58 &	7.46 $\pm$	3.71 &	16.56 $\pm$	9.36 &	19.43 $\pm$	10.69 &	29.89 $\pm$	16.48 \\
850\textunderscore1\textunderscore K &2.02 $\pm$	0.68 &	5.0 $\pm$	1.17 &	7.46 $\pm$	1.75 &	9.55 $\pm$	7.71 &	4.43 $\pm$	4.15 &	5.62 $\pm$	5.84 &	4.61 $\pm$	5.89 \\
850\textunderscore13 &- &	- &	- &	73.78 $\pm$	24.98 &	93.43 $\pm$	28.31 &	130 $\pm$	39 &	78.39 $\pm$	23.33\\
850\textunderscore14 &- &	- &	- &	52.14 $\pm$	14.9 &	78.91 $\pm$	18.42 &	85.72 $\pm$	22.91 &	107 $\pm$	24 \\
850\textunderscore17 &- &	- &	-	 & 91.62 $\pm$	33.41 &	112 $\pm$	27 &	103 $\pm$	25 &	110 $\pm$	25\\
850\textunderscore20 &- &	- &	- &	86.18 $\pm$	27.08 &	137 $\pm$	37 &	129 $\pm$	40 &	76.1 $\pm$	22.79 \\
850\textunderscore22 &- &	- &	- &	31.29 $\pm$	10.45 &	56.36 $\pm$	16.74 &	67.86 $\pm$	23.89 &	140 $\pm$	35\\
850\textunderscore25 &9.25 $\pm$	2.18 &	15.96 $\pm$	3.38 &	18.89 $\pm$	4.05 &	85.41 $\pm$	27.26 &	70.97 $\pm$	27.49 &	105 $\pm$	32 &	79.02 $\pm$	22.71 \\
850\textunderscore31 &2.63 $\pm$	0.81 &	4.32 $\pm$	1.03 &	8.59 $\pm$	1.98 &	19.77 $\pm$	10.58 &	14.54 $\pm$	8.93 & 	40.47 $\pm$	19.89 &	34.77 $\pm$	15.6 \\
850\textunderscore16 &- &- &	- &	84.42 $\pm$	26.66 &	117 $\pm$	35 &	145.42 $\pm$	49.35 &	166 $\pm$	42 \\
850\textunderscore26 &3.08 $\pm$	0.91 &	5.71 $\pm$	1.32 &	11.13 $\pm$	2.49 &	29.33 $\pm$	14.59 &	19.75 $\pm$	12.12 &	36.97 $\pm$	22.32 &	67.18 $\pm$	33.45 \\
850\textunderscore4 &-	 & - & -	 &- &	- &	-	 & -\\
850\textunderscore5 &- &	- &	1.8 $\pm$	0.58 &	- &	- &	- &	- \\
850\textunderscore8 &- &	- &	- &	1.52 $\pm$	1.54 &	7.59 $\pm$	3.91 &	2.43 $\pm$	8.95 &	1.86 $\pm$ 	10.4 \\
850\textunderscore9 &- &	- &	- &	17.08 $\pm$	6.34 &	49.19 $\pm$	14.08 &	63.37 $\pm$	21.07 &	- \\
850\textunderscore24 & - &	- &	- &	31.74 $\pm$	8.33 &	29.72 $\pm$	9.43 &	37.5 $\pm$	13.05 &	-\\
850\textunderscore1\textunderscore J &- &	0.43 $\pm$	0.2 &	0.16 $\pm$	0.14 &	4.8 $\pm$	2.92 &	4.62 $\pm$	2.1 &	8.44 $\pm$	6.71 &	5.4 $\pm$	5.51 \\
\hline
\end{tabular}
\end{table*}

\section{Redshifts For All Non-Cluster Members}

\begin{table}
\caption{Redshifts for the non-cluster members.}  
\label{table:NCRS}     
\small
\centering           
\begin{tabular}{ccccccccccccc} 
\hline\hline   
ID & RA & DEC & $z_{EZ}$ &  $z_{CG}$ \\ \hline
850\textunderscore1\textunderscore I & 222.309 & 8.936 & 2.48$^{+0.72}_{-1.18}$ & 2.41$\pm$1.21\\
850\textunderscore2 & 222.304 & 8.970 & 2.92$^{+0.49}_{-0.54}$& 3.77$\pm$0.74 \\
850\textunderscore3 & 222.225 & 8.897 & 2.59$^{+0.81}_{-1.06}$ & 1.44$\pm$0.56\\
850\textunderscore6 & 222.284 & 8.957 & 2.57$^{+0.33}_{0.34}$ & 3.95$\pm$0.50 \\
850\textunderscore7 & 222.281 & 8.899 & 1.27$^{+0.29}_{-0.30}$ & 1.46$\pm$0.35\\
850\textunderscore10 & 222.340 & 8.878 & 2.40$^{+0.75}_{-0.76}$& 3.01$\pm$0.64\\
850\textunderscore11 & 222.263 & 8.901 & 3.14$^{+0.62}_{-0.61}$ & 3.70$\pm$0.56\\
850\textunderscore15 & 222.380 & 8.968 & 0.94$^{+0.46}_{-0.46}$ & 0.97$\pm$0.30\\
850\textunderscore21 & 222.355 & 8.990 & 2.83$^{+0.67}_{-0.64}$ & 3.25$\pm$0.58\\
850\textunderscore23 & 222.348 & 8.937 & 0.14$^{+0.07}_{-0.09}$& 1.27$\pm$1.31\\
850\textunderscore27 & 222.295 & 8.889 & 2.59$^{+0.43}_{-0.63}$ & 2.43$\pm$0.35\\
850\textunderscore29 & 222.363 & 8.988 & 1.51$^{+0.55}_{-0.47}$ & 1.44$\pm$0.24\\
850\textunderscore32 & 222.317 & 8.902 & 1.46$^{+0.23}_{-0.22}$ & 1.50$\pm$0.18\\
\hline
\end{tabular}
\end{table}


\bsp	
\label{lastpage}
\end{document}